\documentclass{amsart}
\usepackage{amssymb}
\usepackage{amsfonts}
\usepackage{geometry}

\newtheorem{theorem}{Theorem}
\theoremstyle{plain}

\newtheorem{corollary}{Corollary}

\newtheorem{definition}{Definition}

\newtheorem{lemma}{Lemma}

\numberwithin{equation}{section}
\input{tcilatex}

\begin{document}
\title[Strategy-proofness and single peakedness]{Strategy-proofness and
single peakedness in bounded distributive lattices}
\author{Ernesto Savaglio$^{\natural }$ and Stefano Vannucci$^{\ddag }$,}
\address{University of Chieti-Pescara \& GRASS$^{\natural }$, \\
and University of Siena$^{\ddag }$}
\date{May 12, 2014}
\maketitle

\begin{abstract}
Two distinct specifications of single peakedness as currently met in the
relevant literature are singled out and discussed. Then, it is shown that,
under both of those specifications, a voting rule as defined on a bounded
distributive lattice is strategy-proof on the set of all profiles of single
peaked total preorders if and only if it can be represented as an iterated 
\textit{median} of projections and constants, or equivalently as the
behaviour of a certain median tree-automaton. The equivalence of individual
and coalitional strategy-proofness that is known to hold for single peaked
domains in bounded linear orders fails in such a general setting. A related
impossibility result on anonymous coalitionally strategy-proof voting rules
is also obtained.



\end{abstract}

\section{Introduction}

A good decision on an issue of social concern such as location of a public
facility or choice of a tax rate has to rely on some information that is
typically private and disperse among the relevant stakeholders. In many
cases, voting by a suitable committee is one of the most practical means to
elicit and amalgamate such information, and produce the final decision.
Since part of the relevant information is private, however, voters may
attempt to manipulate the outcome by misrepresenting that information (say,
their most preferred outcome), and they are likely to do that \textit{if }%
the voting rule allows for profitable \textit{individual} manipulations.
Now, that manipulative behaviour may easily result in inefficient outcomes,
especially if enacted by several uncoordinated voters. Furthermore, the
perceived availability of profitable manipulations may encourage diversion
of voters' resources to gather private information concerning other voters.
Thus, in order to prevent such possibly wasteful manipulative activities a
voting rule should be reputedly \textit{strategy-proof}, namely immune to
advantageous individual manipulations. If, moreover, voters have access to
cheap communication facilities allowing them to coordinate their voting
strategies, then they can engage in coalitional manipulations as well.
Therefore, in that case a voting rule should also be reputedly \textit{%
coalitionally strategy-proof}, namely immune to jointly profitable
manipulations on the part of \textit{coalitions} of voters.

Those observations raise the following general identification issues:

$\left( i\right) $ \emph{what are (if any) the strategy-proof voting rules
on the relevant preference domain?}

$\left( ii\right) $ \emph{which of them (if any) are also coalitionally
strategy-proof?}

Of course, several alternative domains may be - and have been - taken into
consideration in the aftermath of the Gibbard-Satterthwaite impossibility
theorem. The present paper will address the foregoing issues mainly
focussing on an important class of \textit{single peaked}\footnote{%
`Single peakedness' will be used as a general non-technical term that admits
of several specifications, including of course unimodality and locally
strict unimodality as introduced below.} domains of total preorders in 
\textit{bounded distributive lattices}\footnote{%
A distributive lattice is a partially ordered set such that any two elements 
$x$,$y$ admit a least upper bound or join $x\vee y$ and a greatest lower
bound or meet $x\wedge y$ that mutually `distribute' on each other i.e.
interact much like set-theoretic union and intersection. Clearly, the join
and meet of any finite set of elements are also well-defined.}, namely the
domains denoted here as \textit{unimodal }and \textit{locally strictly
unimodal} (to be defined below). \textit{A characterization of the entire
class of strategy-proof voting rules on the full unimodal and locally
strictly unimodal domains will be provided, generalizing or extending
virtually all previously known results of that kind. }Quite remarkably,%
\textit{\ the simple majority rule or extended median }(that is well-known
to be strategy-proof and coalitionally strategy-proof on both unimodal and
locally strictly unimodal domains in bounded chains)\textit{\ is confirmed
to belong to the strategy-proof class even in the present wider setting. }On
the other hand, \textit{it will also be shown that in a very large class of
bounded distributive lattices that are not linear orders, and under minimal
neutrality\footnote{%
A voting rule is neutral with respect to a certain pair of outcomes when it
treats them in an unbiased manner.} requirements, no anonymous voting rule
is coalitionally strategy-proof on the foregoing domains.}

Single peaked preferences arise in a natural way whenever each agent's
representation of the outcome space is endowed with some `natural' ternary 
\textit{betweenness} relation establishing for any two outcomes $x,y$
whether an arbitrary outcome $z$ lies between $x$ and $y$ or not: indeed,
single peaked total preorders are those total preorders with a \textit{%
unique best outcome }that \textit{respect -are consistent with- such
betweenness relation}.

However, such a broad description of single peakedness is in fact compatible
with \textit{several distinct }specifications of the domain of single peaked
preference relations.

At least two salient issues require further preliminary clarification,
namely:

(a) is the relevant betweenness relation agent-invariant (hence unique) or
agent-dependent, and

(b) what is precisely meant by `consistency of preferences with the relevant
betweenness relations'?

Concerning the first issue, the present paper follows the tradition that can
be traced back at least to Black (1948) and was largely taken for granted in
the early social choice theoretic literature: \textit{the relevant
betweenness relation is required to be agent-invariant hence unique across
voters, }modeling a representation of the outcome structure that is entirely 
\textit{shared }by all the involved parties.

Concerning the second issue, we focus on the two main variants encountered
in the literature on single peakedness, namely

(1) the \textit{`compromise'-view of betweenness-consistency} for
preferences: \textit{if an outcome is intermediate between two outcomes }$x$
and $y$ \textit{then it is to be regarded as a `compromise' between those
two locally `extreme' outcomes and as such is not strictly worse than both }$%
x$ \textit{and} $y$\textit{. Such a `compromise'-view can also be given a
`proximity'-interpretation, relying on a suitable metric induced by the
underlying lattice. That notion of betweenness-consistency is in fact,
arguably, the most `natural' and appropriate one whenever the outcome set is
a distributive lattice; }

(2) an alternative \textit{`top proximity'-view of betweenness-consistency}
for preferences: \textit{if an outcome is intermediate between the top
outcome and another outcome }$y$\textit{\ and distinct from the latter, it
is also closer than }$y$\textit{\ to the top outcome and therefore strictly
better than }$y$. It turns out, however, that \textit{such a notion of `top
proximity' -}when employed in a distributive lattice that is not a chain,
and defined through a suitable latticial metric (as discussed below under
Remark 1)- generates a somewhat spurious notion of local single-peakedness
allowing for the existence of \textit{many local peaks} (more on this point
below).

For the sake of convenience we shall denote as \textit{unimodal }(\textit{%
locally strictly unimodal, }respectively) precisely those preference
profiles of total preorders that are \textit{single peaked under
specification (1) ((2), respectively) }of betweenness-consistency.\textit{\ }%
Indeed, most contributions in the literature on single peakedness and
strategy-proofness of voting rules and social choice functions focus exactly
on unimodal or locally strictly unimodal preference profiles as defined
above.

In a pioneering paper, Moulin (1980) characterizes the class of all
strategy-proof voting rules (or, equivalently, `top-only' social choice
functions) on the domain of all total preorders that are unimodal with
respect to the `natural' betweenness relation of a bounded linearly ordered
outcome set. In fact, he shows that such strategy-proof voting rules are
precisely those based on the median as applied to voters' choices possibly
augmented with a certain set of fixed outcomes aptly dubbed `phantom
vote(r)s' by Border and Jordan (1983). Moreover, in the foregoing work
Moulin points out that \textit{those voting rules are also coalitionally
strategy-proof}.

In a remarkable subsequent contribution, Danilov (1994) provides a similar
median-based characterization of strategy-proof voting rules on the domain
of all linear preference orders that are unimodal with respect to the
`natural' betweenness relation of a (bounded) undirected \textit{tree,} and
establishes \textit{equivalence of individual and coalitional
strategy-proofness} on that domain.

Now, (bounded) linear orders or chains are a quite special subclass of
(bounded) distributive lattices, and many outcome spaces which are of
interest for multi-agent aggregation problems and are not chains do share
with chains precisely that latticial structure. Several examples of such
outcome spaces will be presented and discussed in some detail in Section 3
below. They include spaces consisting respectively of \textit{choice
functions or generalized revealed preference relations, graded evaluations
of items, systems of poverty thresholds, judgments consisting of deductively
closed sets of statements, binary (dis)similarity relations or matrices,
portfolios of basic derivative assets}.

Moreover, a \textit{very natural betweenness relation is available in any
lattice}: just declare $z$ to lie \textit{between }$x$ and $y$ if $z$ is
larger than -or equal to- the meet of $x$ and $y$ (written $x\wedge y$%
\textit{) and }smaller than -or equal to- the join of $x$ and $y$ (written $%
x\vee y$).\footnote{%
That is, for instance, the notion of betweenness underlying `intermediate
preferences' as introduced by Grandmont (1978) and recently reconsidered by
Bossert and Sprumont (2014) in their study of strategy-proof \ preference
aggregation rules. That is also the betweenness relation underlying the
proximity-based interpretation of single peakedness used by Barber\`{a}, Gul
and Stacchetti (1993) who rely on the $L_{1}$-metric (or `taxi-cab' metric).}
Hence, both unimodal and locally strictly unimodal preference domains of
total preorders can be easily defined for all (bounded) distributive
lattices with respect to the standard latticial betweenness, namely: \textit{%
a total preference preorder on a distributive lattice is unimodal if it has
a unique maximum and is `compromise'-consistent (`top proximity-consistent',
respectively) with the latticial betweenness relation.}

No characterizations of strategy-proof voting rules on unimodal domains in 
\textit{general (possibly infinite) bounded distributive lattices other than
chains }are available in the extant literature. In particular, the
equivalence issue concerning strategy-proofness and coalitional
strategy-proofness of voting rules on unimodal domains has never been
addressed before in the foregoing latticial setting.

To be sure, there are some partial results implying existence of
strategy-proof voting rules but not of coalitionally strategy-proof voting
rules on \textit{locally strictly unimodal }domains in some \textit{special}
distributive lattices. But those results \textit{do not address at all the
case of unimodal domains in bounded distributive lattices }since they
variously concern top-proximity-based single peaked preferences with respect
to Euclidean-metric-betweenness in $m$-dimensional Euclidean spaces 
\footnote{%
Recall that Euclidean spaces may be regarded as Riesz spaces i.e. as
partially ordered vector spaces that are also (distributive) lattices.} with 
$m\geq 2$ (see e.g. Border and Jordan (1983), Bordes, Laffond and Le Breton
(2012)), or preference domains such as \textit{locally strictly unimodal
domains in finite products of bounded chains or in finite distributive
lattices }(Barber\`{a}, Gul and Stacchetti (1993) and (Nehring and Puppe
(2007 a, b), respectively) or\textit{\ separable preferences }(Barber\'{a},
Sonnenschein and Zhou (1991))\textit{\ in certain finite Boolean lattices of
cardinality eight or more\footnote{%
A Boolean lattice is a bounded distributive lattice with upper bound $1$ and
lower bound $0$ such that each element $x$ has a \textit{complement} $%
x^{\prime }$ satisfying both $x\vee x^{\prime }=1$ and $x\wedge x^{\prime
}=0 $. }} . The latter preference domains, however, turn out to be \textit{%
disjoint from unimodal domains on that class of lattices. Moreover, the
foregoing works do not address at all the general case of bounded
distributive lattices}.

But then, \textit{what about strategy-proof voting rules on unimodal (or
locally strictly unimodal) domains in arbitrary bounded distributive lattices%
}? Are they \textit{median-representable}? When do they also enjoy \textit{%
equivalence of individual and coalitional strategy-proofness}?

The present paper aims at filling this significant gap in the literature and
provides a study of strategy-proofness and unimodality in\textit{\ general }%
bounded distributive lattices\textit{. }A median-based characterization of
strategy-proof voting rules on unimodal and locally strictly unimodal
domains in bounded distributive lattices is established by introducing 
\textit{median tree-automata representations of voting rules}. It is a
remarkable feature of our characterization that it unifies (generalizing or
extending, and bringing together) several notions, approaches and results
from the extant literature, namely:

\begin{itemize}
\item The characterization contributed by the present paper generalizes
Moulin's original characterization of strategy-proof voting rules on
unimodal domains in bounded chains to both (full) unimodal and locally
strictly unimodal domains in \textit{all} bounded distributive lattices,
obtaining a lattice-polynomial representation which is a dual -and
equivalent- version of that produced by the former author for the special
case of bounded linear orders or chains (Moulin (1980)).

\item Our characterization also highlights the equivalence of that
lattice-polynomial representation to another and new representation of
strategy-proof voting rules on unimodal domains as the \textit{behaviour maps%
} of certain \textit{median tree-automata }acting on suitably labelled
trees. That\textit{\ }tree-automata-theoretic representation essentially
amounts to a streamlining and extension of the approach pioneered by Danilov
(1994) in his remarkable characterization of strategy-proof voting rules on
unimodal domains of linear orders in bounded trees via an
interval-monotonicity property.

\item The lattice-polynomial representation mentioned above is in turn a
generalization of `latticial-federation consensus functions' or,
equivalently, of `generalized committee voting rules' as introduced
respectively, and independently, by Monjardet (1990) in his path-breaking
contribution to (non-strategic) aggregation problems in latticial
structures, and by Barber\'{a}, Sonnenschein and Zhou (1991) in their
well-known study of strategy-proof voting mechanisms on \textit{separable}
preference domains in finite Boolean lattices.

\item Finally, it is also proved that the equivalence between
strategy-proofness and coalitional strategy-proofness - that is known to
hold for both unimodal and locally strictly unimodal domains in bounded
linear orders and for unimodal domains of linear orders in bounded trees- 
\textit{fails} for both unimodal and locally strictly unimodal domains in
bounded distributive lattices that are \textit{not} linear orders,\footnote{%
Vannucci (2012) provides a general incidence-geometric argument to explain
that equivalence-failure.} hence \textit{even in outcome spaces with a
well-defined (and unique) median operation }(Theorem 2). An impossibility
theorem concerning coalitional strategy-proofness on the full unimodal and
locally strictly unimodal domains for anonymous voting rules satisfying very
weak local sovereignty and neutrality requirements (Theorem 3) is also
provided. Thus, in particular, Theorem 3 establishes that the former
equivalence may fail in bounded distributive lattices \textit{even for
non-sovereign voting rules}\footnote{%
Indeed, Nehring and Puppe (2007 (b)) prove that the only efficient and
strategy-proof voting rules on certain `rich' subdomains of locally strictly
unimodal domains of linear orders in finite Boolean lattices or $m$%
-hypercubes $\mathbf{2}^{m\text{ }}$with $m\geq 3$ are (weakly) dictatorial.
Notice, however, that such preference domain is incomparable to our unimodal
domain and (weakly) efficient voting and social choice rules are in
particular sovereign: thus, Nehring and Puppe's result pertains to a class
of rules which is utterly non-comparable to the class of voting rules
covered by Theorem 3 of the present work.}\textit{.}
\end{itemize}

The remainder of the paper is organized as follows. The next section
describes several remarkable examples of bounded distributive lattices that
occur in some well-known aggregation problems. Section 3 introduces the
notation and definitions and includes the main results of the paper on the
structure of the strategy-proof voting rules for full unimodal and locally
strictly unimodal domains of total preorders in arbitrary bounded
distributive lattices. In Section 4 the main results of the present work are
discussed in some detail with \ reference to a simple example concerning the
Boolean square. Section 5 includes a detailed discussion of some related
literature and offers some concluding remarks. Appendix 1 collects all the
proofs. Appendix 2 is devoted to a detailed presentation of the basic
notions on tree automata used in the paper.

\section{What kind(s) of single peakedness and why bounded distributive
lattices?}

Thus, our analysis shall be focussed on strategy-proof voting rules for both 
\textit{unimodal and locally strictly unimodal }preference profiles in%
\textit{\ }bounded distributive lattices as informally defined in the
Introduction. Therefore, the model \ to be introduced below applies under
the following conditions:

\bigskip

(1) the outcome set is a partially ordered set $\mathcal{X}=\left(
X,\leqslant \right) $ with a top and a bottom, and such that the (binary)
least-upper bound $\vee $ and greatest-lower bound $\wedge $ as induced by $%
\leqslant $ satisfy the distributive identity $x\wedge (y\vee z)=(x\wedge
y)\vee (x\wedge z)$ for any $x,y,z\in X$;

(2) all voters are prepared to assess outcomes according to the latticial
ternary betweenness relation of $\mathcal{X}$ denoted $B_{\mathcal{X}}$ :
namely, outcome $z$ lies between outcomes $x$ and $y$ if and only if $%
x\wedge y\leqslant z\leqslant x\vee y$ ;

(3) the preferences of all voters are consistent with the latticial
betweenness relation in one of the two following senses: (i) (unimodality)
any outcome that $z$ lies between outcomes $x$ and $y$ is to be regarded as
a `compromise' between its relative extrema $x$ and $y$ and is therefore not
strictly worse than each one of the latter; (ii) (locally strict
unimodality) any outcome that $z$ lies between top outcome $x$ and outcome $%
y $ and is distinct from $y$ is to be regarded as `closer' to the top
outcome than $y$ and is therefore strictly better than $y$.\bigskip

Observe that when specialized to the particular case of a bounded chain,
conditions (1)-(2)-(3(i)) (or (1)-(2)-(3(ii))) are a list of plausible
requirements to be met in order to justify single peaked domains in the
standard version pioneered by Black (1948). Indeed, the foregoing conditions
-especially (1)-(2)-(3(i))- arguably provide the `right' extension of the
notion of a single peaked preference \textit{profile }for distributive
lattices as illustrated in the simple example concerning Boolean squares as
discussed above. Of course, even if condition (1) obtains, conditions (2)
and/or (3) may or may not hold depending on the problem under scrutiny. But
it is at least conceivable that there are interesting cases where conditions
(1)-(2) and either (3(i)) or (3(ii)) are jointly satisfied, and when that is
the case, focusing on the full unimodal (or locally strictly unimodal)
preference domain of a bounded distributive lattice as defined in the
present work seems to be fully justified, indeed somewhat compelling.

In order to fully appreciate the remarkably wide scope and relevance of the
proposed setting let us consider just a few prominent classes of examples of
bounded distributive lattices of special interest, namely:\bigskip

\textbf{Example 1}: \textbf{Committee decision on multidimensional binary
issues: aggregation of points on a finite Boolean hypercube.}

Let $\mathbf{2}^{k}$ be the set of points of a finite $k$-dimensional
Boolean hypercube and $\leq $ the standard componentwise order. Then, take
the $\mathcal{X=}\left( \mathbf{2}^{k},\leq \right) $. When\ considering an
abstract location problem on that discrete cube such as committee-selection
of the \textit{appropriate profile of binary criteria to be satisfied by
candidates}\ in order to qualify for a certain position: here, one has to
face the issue of aggregating the alternative proposals (namely points of $%
\mathbf{2}^{k}$) advanced by members of a panel committee.\bigskip

\textbf{Example 2}: \textbf{Committee selection of location on a
multidimensional box: aggregation of points in a product of bounded subsets
of the extended real line}.

Let $\mathbb{R}_{\ast }=\mathbb{R}\cup \left\{ -\infty ,+\infty \right\} $
denote the extended real line, $\leq ^{\ast }$the component-wise extended
natural order on $\mathbb{R}_{\ast }^{m}$, $Y_{i}\subseteq \mathbb{R}_{\ast
} $ for each $i=1,...,m$, and $x,y\in \dprod\limits_{i=1}^{m}Y_{i}$ with $%
x\leq ^{\ast }y$. Then, take $\mathcal{X=}(X,\leqslant )$ with $X=\left\{
z\in \dprod\limits_{i=1}^{m}Y_{i}:x\leq ^{\ast }z\leq ^{\ast }y\right\} $
and $\leqslant =\leq _{|X}^{\ast }$(recall that a product of distributive
lattices is a distributive lattice under the component-wise order). This is
the setting of Barber\`{a}, Gul and Stacchetti (1993) study of
strategy-proofness on locally strictly unimodal domains. If $m=1$, $%
(X\leqslant )$ reduces to a bounded chain, which gives the original standard
setting of the literature on strategy-proofness on single peaked domains,
including the seminal work of Moulin (1980) on the characterization of
strategy-proof voting rules on unimodal domains in a bounded real chain,
where $\mathcal{X=}(\mathbb{R}_{\ast },\leq ^{\ast })$.\medskip

\textbf{Example 3}: \textbf{Election of a representative body and committee
selection of the target of an `atomic' package bid in a combinatorial
auction: aggregation of subsets of a fixed set}.

Let $Y$ be a set of items, $\mathcal{P}(Y)$ its power set and $\Sigma $ $%
\mathcal{\subseteq P}(Y)$ a field of sets (namely $\Sigma $ is nonempty and
such that $\left\{ A\cap B,A\cup B,X\smallsetminus A\right\} \subseteq
\Sigma $ for any $A,B\in \Sigma $). Then, take $\mathcal{X=}(\Sigma
,\subseteq )$. This kind of domain arises in a most natural way in a few
cases including \textit{combinatorial} social choice problems, i.e. social
choice issues concerning \textit{mutually compatible }objects\textit{\ }%
(e.g. selection by committee decision of a representative body, or of the
target of an admissible package bid in a combinatorial auction namely the
item-subset to bid for: in the latter case, $\Sigma $ denotes the set of
admissible packages as fixed by the auction mechanism designer with a view
to keep communication complexity under some acceptable threshold).\bigskip

\textbf{Example 4}:\textbf{\ Committee selection of a portfolio of basic
derivative assets: aggregation of points in a bounded Riesz space.}

Let $s:[0,1]\rightarrow \mathbb{R}_{+}$ be a continuous non-negative
real-valued function denoting a limited-liability state-dependent stock with
state-space $[0,1]$, and $b:$ $[0,1]\rightarrow \mathbb{R}_{+}$ a constant
function denoting the relevant bond; if $s([0,1])=$ $[\alpha ,\beta ]$ then
ordered linear space $(X=\mathcal{C[}\alpha ,\beta ],\leqslant )$, the set
of continuous real-valued functions on $\left[ \alpha ,\beta \right] $
endowed with the component-wise natural order, denotes the space of all 
\textit{continuous options on }$s$. It turns out that $(\mathcal{C[}\alpha
,\beta ],\leqslant )$ is in fact a \textit{Riesz space namely it is also a
(distributive) lattice:} moreover, it is \textit{bounded }with constant
functions $f_{\alpha }$ and $f_{\beta }$ as bottom and top elements,
respectively. In particular, the latticial operations $\vee $ and $\wedge $
of $(\mathcal{C[}\alpha ,\beta ],\leqslant )$ enable convenient
representations of both call options on $s$ and put options on $s$ at any
striking price $p$ as $(s-pb)\vee 0$ and $(pb-s)\vee 0$, respectively, and
it can be shown that each continuous option in $\mathcal{C[}\alpha ,\beta ]$
can be represented as\textit{\ a portfolio of call options }(see e.g. Brown
and Ross (1991)). Thus, under the foregoing stipulations a committee
selection of a continuous option on a stock (or equivalently of a portfolio
of call options on that stock) amounts to an aggregation of points in
bounded Riesz space $\mathcal{X=}(\mathcal{C[}\alpha ,\beta ],\leqslant )$%
.\bigskip

\textbf{Example 5: General revealed preference aggregation: aggregation of
choice functions on a fixed set}.

Let $Y$ be a set of items, and $\mathcal{P}(Y)$ its power set. A
(full-domain) \textit{choice function }on $Y$ is a function $f:\mathcal{P}%
(Y)\rightarrow \mathcal{P}(Y)$ such that $f(A)\subseteq A$ for each $%
A\subseteq Y$. Now, denote by $\mathcal{C}_{Y}$ the set of all choice
functions on $Y$, and for any $f,g\in $ $\mathcal{C}_{Y}$ posit $f\leqslant
^{\prime }g$ if and only if $f(A)\subseteq g(A)$ for all $A\subseteq Y$.
Then, take $\mathcal{X=}(\mathcal{C}_{Y},\leqslant ^{\prime })$, where the
constant empty-valued choice function is the bottom and the identity choice
function is the top. The aggregation of the choice functions in $\mathcal{C}%
_{Y}$ may be regarded as a \textit{natural generalization of the classic
problem of preference aggregation in social welfare analysis} if preferences
are taken to summarize choice behaviour and the usual `consistency'
requirements related to acyclicity properties are relaxed. Indeed, the issue
here is the elicitation and aggregation of complete lists of recommendations
concerning local choice behaviour from a population of experts and/or
stakeholders.\bigskip

\textbf{Example 6}: \textbf{Merging databases of binary (dis)similarity
coefficients: aggregation of dissimilarity and tolerance relations on a
fixed set}.

Let $Y$ a set of items: a \textit{dissimilarity} (or orthogonality) relation
on $Y$ is an irreflexive and symmetric binary relation $D$ on $Y$ i.e. $%
D\subseteq Y\times Y$ is such that (i) $(y,y)\notin D$ for all $y\in Y$ and
(ii) $(y,z)\in D$ implies $(z,y)\in D$ for all $y,z\in Y$. Denote by $%
\mathcal{D}_{Y}$ the set of all dissimilarity relations on $Y$, and take $%
\mathcal{X=}(\mathcal{D}_{Y},\subseteq )$; a \textit{tolerance} (or
similarity) relation on $Y$ is a reflexive and symmetric binary relation $D$
on $Y$ i.e. $D\subseteq Y\times Y$ is such that (i) $(y,y)\in D$ for all $%
y\in Y$ and (ii) $(y,z)\in D$ implies $(z,y)\in D$ for all $y,z\in Y$.
Denote by $\mathcal{T}_{Y}$ the set of all tolerance relations on $Y$, and
take $\mathcal{X=}(\mathcal{T}_{Y},\subseteq )$. Dissimilarity and tolerance
relations are one of the basic inputs in most algorithmic classification
procedures: if many (binary) dissimilarity databases from several distinct
sources are available, one may wish to aggregate them to produce a \textit{%
unique consensus database.}\bigskip

\textbf{Example 7}: \textbf{Merging judgments with their implications:
aggregation of order filters over a partially ordered set}.

Let $\mathcal{Y=}(Y,\leqslant )$ denote a finite partially ordered
population. An \textit{order filter }of $\mathcal{Y}$ is a set $F\subseteq Y$
such that for all $y,z\in Y$, $z\in F$ whenever $y\in F$ and $y\leqslant z$.
Denote by $\mathcal{F}_{\mathcal{Y}}$ the set of all order filters of $%
\mathcal{Y}$, and take $\mathcal{X=}(\mathcal{F}_{\mathcal{Y}},\subseteq )$.
Order filters may variously arise in several aggregation problems, including 
\textit{judgment aggregation problems with implication-constrained agendas}.
Thus, in the latter case $Y$ denotes a collection of propositions (namely,
sets of logically equivalent sentences) and $\leqslant $ denotes the
relevant \textit{implication} or consequence relation between propositions.
In that connection, a judgment amounts to a deductively closed set of
propositions, namely an order filter of $\mathcal{Y}$, and establishing a 
\textit{consensus judgment }reduces to aggregating order filters\ of $%
\mathcal{Y}$.\bigskip

\textbf{Example 8}: \textbf{Merging proposals for multidimensional poverty
thresholds: aggregation of order ideals over a partially ordered set}.

An \textit{order ideal }of a partially ordered set $\mathcal{Y=}$ $%
(Y,\leqslant )$ is a set $I\subseteq Y$ such that for all $y,z\in Y$, $z\in
I $ whenever $y\in I$ and $z\leqslant y$. Denote by $\mathcal{I}_{\mathcal{Y}%
}$ the set of all order ideals of $\mathcal{Y}$, and take $\mathcal{X=}(%
\mathcal{I}_{\mathcal{Y}},\subseteq )$. Order ideals are also relevant to
several aggregation problems, including choice of a (system of)\textit{\
threshold}(s) in \textit{multidimensional poverty analysis}: a list of
relevant binary attributes is considered, and different thresholds namely
combinations of minimal deprivations are proposed by qualified experts
and/or political representatives to identify the poor. Each threshold
corresponds to an order ideal, hence amalgamating the advanced proposals
amounts to aggregating order ideals.\bigskip

\textbf{Example 9}: \textbf{Merging graded assessments, and computing
reputations: aggregation of graded evaluations}.

Let $\Lambda \mathcal{=}(L,\leq )$ denote a (bounded) linearly ordered set
of grades, $X$ a (finite) population of candidates to be evaluated, and $N$
a (finite) population of evaluators. Then, denote by $L^{X}$ the set of all
possible gradings of $X$, by $\leqslant $ the point-wise partial order
induced by $\leq $, and take $\mathcal{X=}(L^{X},\leqslant )$. This is
indeed the formal setting recently proposed by Balinski and Laraki (2010) in
order to advance their case for \textit{majority judgment}. It may be
considered for aggregating grades achieved by a population of students in
different subjects, assessments of wines according to several alternative
graded criteria or the graded performances of participants in a multi-trial
competition. A particular case of special interest is provided by the
definition of reputation systems, both off-line and on-line. Indeed, by
taking $L\subseteq \mathbb{Z}$ (or $L\subseteq $ $\mathbb{Q}$) with the
natural order of the integers (or the rationals), and $N=X$, a point of $%
L^{N}$ denotes the approval/citation profile of an author (or the on-line
feedback profile of a web-node), and a \textit{reputation system }is an
aggregation function $f:(L^{N})^{N}\rightarrow L^{N}$.\bigskip

The foregoing set of examples is of course not meant to be an exhaustive
list, and some of them may well refer to comparatively more uncommon or
hypothetical decision problems than others. However, that list provides in
our view a quite representative sample of the wide class of interesting
aggregation problems to which our results on strategy-proof voting rules in
bounded distributive lattices do in fact apply.

\section{Model and results}

Let $N=\left\{ 1,...,n\right\} $ denote the finite population of voters, and 
$\mathcal{X=}(X,\leqslant )$ the partially ordered set of alternative
outcomes (i.e. $\leqslant $ is a reflexive, transitive and antisymmetric
binary relation on $X$). We suppose $\left\vert N\right\vert \geq 3$ in
order to avoid tedious qualifications, and denote as $x||y$ any pair of $%
\leqslant $-incomparable outcomes.\footnote{%
We denote by $\left\vert \cdot \right\vert $ the cardinality of a set.} Let
us also assume that $\mathcal{X}=(X,\leqslant )$ is a \textbf{distributive
lattice}\textit{\ }namely both the \textit{least-upper-bound (l.u.b.)} $%
\wedge $ and the \textit{greatest-lower-bound} \textit{(g.l.b.) }$\vee $ of
any $x,y\in X$ as induced by $\leqslant $ are well-defined binay operations
on $X$, and\textit{\ }for all $x,y,z\in X$, $x\wedge (y\vee z)=(x\wedge
y)\vee (x\wedge z)$ (or, equivalently, $x\vee (y\wedge z)=(x\vee y)\wedge
(x\vee z)$).\footnote{%
Notice that thanks to associativity of $\vee $ and $\wedge $ the l.u.b. and
the g.l.b. of \textit{any finite} $Y\subseteq X$ are also well-defined and
denoted by $\vee Y$ \ and $\ \wedge Y$, respectively; if $Y$ is infinite $%
\vee Y$ \ and $\ \wedge Y$ may or may not be well-defined.} In particular, $%
\mathcal{X}=(X,\leqslant )$ is a \textbf{linear order }or \textbf{chain }if [%
$x\leqslant y$ or $y\leqslant x$] holds for all $x,y\in X$ (recall that, as
it is easily checked, a chain does indeed satisfy the distributive identity
above). A \textit{join irreducible }element of $\mathcal{X}$ is any $j\in X$
such that $j\neq \wedge X$ and for any $Y\subseteq X$ if $j=\vee Y$ then $%
j\in Y$. The set of all join irreducible elements of $\mathcal{X}$ is
denoted $J_{\mathcal{X}}$. An \textit{atom }of a lower bounded $\mathcal{X}$
is any $\leqslant $-minimal $x\in X\smallsetminus \left\{ \bot \right\} $.
The set of all atoms of $\mathcal{X}$ is denoted $A_{\mathcal{X}}$: clearly, 
$A_{\mathcal{X}}\subseteq J_{\mathcal{X}}$. Moreover, a (distributive)
lattice $\mathcal{X}$ is said to be \textbf{lower (upper) bounded} if there
exists $\bot \in X$ \ ($\top \in X)$ such that $\bot \leqslant x$ $\ $($%
x\leqslant \top $) for all $x\in X$, and \textbf{bounded }if it is both
lower bounded and upper bounded. A bounded distributive lattice $%
(X,\leqslant )$ is \textbf{Boolean }if for each $x\in X$ there exists a 
\textit{complement }namely an $x^{\prime }\in X$ such that $x\vee x^{\prime
}=\top $ and $x\wedge x^{\prime }=\bot $. A ternary \textbf{betweenness }%
relation $B_{\mathcal{X}}=\left\{ (x,z,y)\in X^{3}:x\wedge y\leqslant
z\leqslant x\vee y\right\} $ is defined on $\mathcal{X}$, and $x,y\in X$, $%
[x,y]=\left\{ z\in X:x\wedge y\leqslant z\leqslant x\vee y\right\} $ is the 
\textit{interval }induced by $x$ and $y:$ therefore, \textit{for any} $%
x,y,z\in X,$ $z\in \lbrack x,y]$ \textit{if and only if} $(x,z,y)\in B_{%
\mathcal{X}}$ (\textit{also written }$B_{\mathcal{X}}(x,z,y)$).\footnote{%
The ensuing analysis could be pursued by replacing entirely betweenness
relations with intervals (see Vannucci (2012) for such an approach in a more
general setting).}

It is a remarkable fact that a ternary operation called \textit{median} is
well-defined on an arbitrary distributive lattice.

\begin{definition}
The \textbf{median}\textit{\ }on $\mathcal{X}$ is the ternary operation $\mu
:X^{3}\rightarrow X$ defined as follows: for all $x,y,z\in X$,%
\begin{equation*}
\mu (x,y,z)=(x\wedge y)\vee (y\wedge z)\vee (x\wedge z)\text{.}
\end{equation*}
\end{definition}

Notice that, due to commutativity and associativity of $\wedge $ and $\vee $
the median $\mu $ as defined above is \textit{invariant under permutations
of its arguments }or \textit{symmetric,} namely for any $x_{1},x_{2},x_{3}%
\in X$ and any permutation $\sigma :\left\{ 1,2,3\right\} \longrightarrow
\left\{ 1,2,3\right\} $, $\mu (x_{\sigma (1)},x_{\sigma (2)},x_{\sigma
(3)})=\mu (x_{1},x_{2},x_{3}).$\footnote{%
It should be recalled here that Birkhoff and Kiss (1947) also provide a
general characterization of the median in a (bounded) distributive lattice
through the following axioms for an arbitrary ternary operation $m$ on a set 
$A:$%
\par
m(i) there exist $0,1\in A$ such that $m(0,a,1)=a$ for all $a\in A$;
\par
m(ii) $m(a,b,a)=a$ for all $a,b\in A$:
\par
m(iii) $m(a,b,c)=m(b,a,c)=m(b,c,a)$ \ for all $a,b,c\in A$;
\par
m(iv) $m(m(a,b,c),d,e)=m(m(a,d,e),b,m(c,d,e))$ for all $a,b,c,d,e\in A$.
\par
An alternative characterization of the latticial median is provided by
Sholander (1954(a)).}\medskip

\textbf{Remark 1.}Notice that a natural median-based betweenness relation $%
B_{\mathcal{X}}^{\mu }\subseteq X^{3}$ can be defined on $\mathcal{X}$ by
the following rule: for any $x,y,z\in X$, \ $(x,z,y)\in B_{\mathcal{X}}^{\mu
}$ iff $\mu (x,y,z)=z$. But it is easily shown that in fact $B_{\mathcal{X}%
}^{\mu }=B_{\mathcal{X}}$ \ (see\ Birkhoff and Kiss (1947), Theorem 1).
Moreover, if the relevant distributive lattice $(X,\leqslant )$ is metric
i.e. is endowed with a positive valuation namely a function $v:X\rightarrow 
\mathbb{R}$ such that $v(x\vee y)=v(x)+v(y)-v(x\wedge y)$ and $v(x)<v(y)$
whenever $x<y$, then it can be shown that $B_{\mathcal{X}}=\left\{
(x,y,z)\in X^{3}:d_{v}(x,z)=d_{v}(x,y)+d_{v}(y,z)\right\} $, where $d_{v}$
is the metric induced by $v$ as defined by the rule $d_{v}(x,y)=v(x\vee
y)-v(x\wedge y)$ \ (see Glivenko (1936), Theorem V). Thus $B_{\mathcal{X}%
}(x,y,z)$ holds precisely when $y$ lies on a $d_{v}$-geodesics or $d_{v}$%
-shortest path joining $x$ and $z$. That fact suggests the possibility to
provide $B_{\mathcal{X}}$ with a straightforward metric representation
whenever deemed appropriate, and highlights the focal role of $B_{\mathcal{X}%
}$ and of the median operation for any plausible proximity relation
respecting the latticial structure of $\mathcal{X}$.\medskip

A few remarkable basic properties of $B_{\mathcal{X}}$ are listed
below:\medskip

\textbf{Claim 1. }The latticial betweenness relation $B_{\mathcal{X}}$
satisfies the following conditions:

$(i)$ symmetry: for all $x,y,z\in X$, if $\ B_{\mathcal{X}}(x,z,y)$ then $B_{%
\mathcal{X}}(y,z,x)$;

$(ii)$ closure (or reflexivity): for all $x,y\in X,$ $B_{\mathcal{X}}(x,x,y)$
and $B_{\mathcal{X}}(x,y,y)$;

$(iii)$ idempotence: for all $x,y\in X$, $B_{\mathcal{X}}(x,y,x)$ only if $%
y=x$;

$(iv)$ convexity (or transitivity): for all $x,y,z,u,v\in X$, if $B_{%
\mathcal{X}}(x,u,y),$ $B_{\mathcal{X}}(x,v,y)$ and $B_{\mathcal{X}}(u,z,v)$
then $B_{\mathcal{X}}(x,z,y)$;

$(v)$ antisymmetry: for all $x,y,z\in X$, if $B_{\mathcal{X}}(x,y,z)$ and $%
B_{\mathcal{X}}(y,x,z)$ then $x=y$.\medskip

Now, consider the set $T_{X}$ of all \textit{topped} total preorders on $X$
(i.e. \textit{connected}, reflexive, and transitive binary relations having
a unique maximum in $X$). For any $\succcurlyeq \in T_{X}$, $%
top(\succcurlyeq )$ denotes the unique maximum of $\succcurlyeq $ (while $%
\succ $ and $\sim $ denote the asymmetric and symmetric components of $%
\succcurlyeq $, respectively).

\begin{definition}
A topped total preorder $\succcurlyeq \in T_{X}$ is \textbf{unimodal}\textit{%
\ }(with respect to $B_{\mathcal{X}}$) if and only if, for each $x,y,z\in X$%
, $z\in \left[ x,y\right] $ implies that either $z\succcurlyeq x$ or $%
z\succcurlyeq y$ (or both).
\end{definition}

As mentioned above, the rationale underlying single peakedness as \textit{%
unimodality }may be plainly described as follows: an unimodal total
preference preorder \textit{respects }betweenness $B_{\mathcal{X}}$ in that
it never regards an intermediate or compromise outcome as strictly worse
than both of its `extreme'-generators.

An alternative notion of single peakedness has also been widely adopted in
the literature under several labels including `generalized single
peakedness'(see e.g. Nehring and Puppe (2007 (a,b) among others). It will be
relabeled here `\textit{locally strict unimodality}' for the sake of
convenience, and may be formulated as follows in the present setting:

\begin{definition}
A topped total preorder $\succcurlyeq \in T_{X}$ (with top outcome $x^{\ast
} $) is \textbf{locally strictly unimodal}\textit{\ }(with respect to $B_{%
\mathcal{X}}$) if and only if, for each $y,z\in X$, $z\in \left[ x^{\ast },y%
\right] \smallsetminus \left\{ y\right\} $ implies $z\succ y$.\medskip
\end{definition}

\textbf{Remark 2.} It is worth noticing here that in the extant literature
unimodality and locally strict unimodality are not always firmly
distinguished as they should be. For instance, in a very interesting and
widely cited paper Nehring and Puppe (2007 (b), p.135) quote Moulin (1980)
as a contribution on `generalized single peaked'(i.e. locally strictly
unimodal) preferences in the case of a line (but see also Barber\`{a}, Gul
and Stacchetti (1993) who identify single peakedness and locally strict
unimodality, suggesting that this is precisely the notion underlying
Moulin's work). However, Moulin's definition, once reformulated in terms of
preferences (as opposed to utilities, as in the original Moulin (1980), p.
439) amounts to the following requirement: `If $a$ is the top outcome or
peak on the line $(X,\leqslant )$ then $a\succ x\succcurlyeq y$ if $%
a<x\leqslant y$ or $y\leqslant x<a$'. Notice however that this condition is
only consistent with \textit{unimodality }as opposed to locally strict
unimodality. To see this just consider $X=\left\{ a,x,y\right\} $ with $%
a<x<y $, and total preorder $\succcurlyeq $ such that $a\succ x\sim y$: \ by
construction, $\succcurlyeq $ is certainly consistent with Moulin's
condition, and it is in fact unimodal but \textbf{not} at all locally
strictly unimodal (or `generalized single peaked').\medskip

By definition, existence of unimodal total preorders with the respect to
latticial betweenness $B_{\mathcal{X}}$ on a lattice $\mathcal{X=(}%
X,\leqslant )$ is clearly not an issue: for any $x\in X$, the total preorder
with $x$ as its unique top outcome and $y\sim z$\ for any other $y,z\in X$
is by construction unimodal. On the other hand, existence of locally
strictly unimodal total preorders with respect to $B_{\mathcal{X}}$ requires
a more detailed argument that relies on some specific properties of $B_{%
\mathcal{X}}$ as combined with the following notions of \textit{%
Suzumura-consistency }(henceforth, \textit{S-consistency}) and \textit{%
non-trivial total extension }of a binary relation:

\begin{definition}
(a) A binary relation $\succcurlyeq $ on $X$ is \textbf{S-consistent}
whenever for all $x,y\in X$, if for some positive integer $k$ there exist $%
z_{1},...,z_{k}\in X$ such that $x\succcurlyeq z_{1}$, $z_{k}\succcurlyeq y$
and $z_{h}\succcurlyeq z_{h+1}$, $h=1,...,k-1$ then \textbf{not }$y\succ x;$

(b) A binary relation $\succcurlyeq ^{\prime }$ is a \textbf{non-trivial
extension }of binary relation $\succcurlyeq $ on $X$ if for all $x,y\in X$:
(i) $x\succcurlyeq y$ entails $x\succcurlyeq ^{\prime }y$; (ii) $x\succ y$
entails $x\succ ^{\prime }y$.
\end{definition}

The former notions are mutually related thanks to a generalization of
Szpilrajn' theorem on ordering extensions due to Suzumura, namely:\bigskip

\textbf{Suzumura's Theorem }(see e.g. Bossert and Suzumura (2010), Theorem
2.8) \emph{A binary relation }$\succcurlyeq $\emph{\ on a set }$X$\emph{\
admits a total preorder as a non-trivial extension if and only if it is
S-consistent.}\bigskip

It turns out that the following claim can be quite easily
established:\bigskip

\textbf{Claim 2.} Let $\mathcal{X=}(X,\leqslant )$ be a (bounded)
distributive lattice, $B_{\mathcal{X}}$ its (latticial) betweenness relation
as defined above, $x\in X$ and $\succ _{x}$ the binary relation on $X$
defined as follows: for all $x,y,z\in X$, $y\succ _{x}z$ if and only if $\
B_{\mathcal{X}}(x,y,z)$ and $y\neq z$ (i.e. $y\in \lbrack x,z]\smallsetminus
\left\{ z\right\} $). Then $\succ _{x}$ admits a non-trivial extension $%
\succcurlyeq _{x}^{\ast }$which is a locally strictly unimodal total
preorder on $X$ with respect to $B_{\mathcal{X}}$.\bigskip

Let $U_{\mathcal{X}}\subseteq T_{X}$ denote the \emph{set of all unimodal
total preorders} (with respect to $B_{\mathcal{X}})$, and $U_{\mathcal{X}%
}^{N}$ the set of all $N$-\textit{profiles }of unimodal total preorders or
full unimodal domain (with respect to $B_{\mathcal{X}}$). Similarly, $S_{%
\mathcal{X}}\subseteq T_{X}$ is the \emph{set of all locally strictly
unimodal total preorders} (with respect to $B_{\mathcal{X}})$, and $S_{%
\mathcal{X}}^{N}$ \ denotes the set of all $N$-\textit{profiles }of locally
strictly unimodal total preorders or full locally strictly unimodal domain
(with respect to $B_{\mathcal{X}}$).

A \textit{voting rule }for $(N,X)$ is a function $f:X^{N}\rightarrow X$. For
any profile $(Y_{i})_{i\in N}$ (where $Y_{i}\subseteq X$ for all $i\in N$) a 
\textit{restricted voting rule }for $(N,X)$ is a function $f:\Pi _{i\in
N}Y_{i}\rightarrow X$. The following properties of a voting rule will play a
crucial role in the ensuing analysis:

\begin{definition}
A voting rule $f:\Pi _{i\in N}Y_{i}\rightarrow X$ is $B_{\mathcal{X}}$-%
\textbf{monotonic }if and only if for all $x_{N}=(x_{j})_{j\in N}\in Y^{N}$, 
$i\in N$ and $x_{i}^{\prime }\in Y$: $f(x_{N})\in \lbrack
x_{i},f(x_{i}^{\prime },x_{N\smallsetminus \left \{ i\right \} })]$.
\end{definition}

\begin{definition}
For any $i\in N$, let $D_{i}\subseteq U_{\mathcal{X}}$ such that $%
top(\succcurlyeq )\in Y_{i}$ for all $\succcurlyeq \in D_{i}$. Then, $f:\Pi
_{i\in N}Y_{i}\rightarrow X$ is (individually) \textbf{strategy-proof}%
\textit{\ }on $\Pi _{i\in N}D_{i}\subseteq U_{\mathcal{X}}^{N}$ if and only
if, for all $x_{N}\in $ $\Pi _{i\in N}Y_{i}$ , $i\in N$ and $x^{\prime }\in
Y_{i}$, and for all $\mathbf{\succcurlyeq }=(\succcurlyeq _{j})_{j\in N}\in
\Pi _{i\in N}D_{i}$, $f(top(\mathbf{\succcurlyeq }_{i}),x_{N\smallsetminus
\left \{ i\right \} })\succcurlyeq _{i}$ $f(x^{\prime },x_{N\smallsetminus
\left \{ i\right \} })$.
\end{definition}

\begin{definition}
For any $i\in N$, let $D_{i}\subseteq U_{\mathcal{X}}$ such that $%
top(\succcurlyeq )\in Y_{i}$ for all $\succcurlyeq \in D_{i}$. Then, $f:\Pi
_{i\in N}Y_{i}\rightarrow X$ is \textbf{coalitionally strategy-proof}\textit{%
\ }on $\Pi _{i\in N}D_{i}\subseteq U_{\mathcal{X}}^{N}$ if and only if for
all $x_{N}\in $ $\Pi _{i\in N}Y_{i}$ , $C\subseteq N$ and $x_{C}^{\prime
}\in \Pi _{i\in C}Y_{i}$, and for all $\mathbf{\succcurlyeq }=(\succcurlyeq
_{j})_{j\in N}\in \Pi _{i\in N}D_{i}$, there exists $i\in C$ such that $%
f(x_{N})\succcurlyeq _{i}$ $f(x_{C}^{\prime },x_{N\smallsetminus C})$.
\end{definition}

The following properties of voting rules will also be considered in the
ensuing analysis.

A voting rule $f:\Pi _{i\in N}Y_{i}\rightarrow X$ is (weakly) \textbf{%
efficient}\textit{\ }if and only if for all $(\succcurlyeq _{j})_{j\in N}\in 
$ $\Pi _{i\in N}D_{i}\subseteq U_{\mathcal{X}}^{N}$ and $y\in X$, $y\notin
f((top(\succcurlyeq _{j})_{j\in N}))$ if there exists $x\in X$ such that $%
x\succ _{j}y$ for all $j\in N$, \textbf{anonymous }if $f((x_{j})_{j\in
N}))=f((x_{\sigma (j)})_{j\in N}))$ for all $x_{N}\in X^{N}$ and all
permutations $\sigma :N\rightarrow N$, \textbf{locally JI-neutral }on $%
Y\subseteq X$ if $\ f((\tau _{jk}(x_{i}))_{i\in N})=\tau
_{jk}(f((x_{i})_{i\in N}))$ for all $y_{N}\in Y^{N}$ and $j,k\in J_{\mathcal{%
X}}\cap Y$ (where $\tau _{jk}:Y\rightarrow Y$ is the elementary permutation
of $Y$ such that $\tau _{jk}(j)=k,$ $\tau _{jk}(k)=j$ and $\tau _{jk}(x)=x$
for any $x\neq j,k$), \textbf{locally sovereign }on $Y\subseteq X$ $\ $if
for all $z\in Y$ there exists $y_{N}\in Y^{N}$ such that $f(y_{N})=z$\textbf{%
, }\ and \textbf{locally idempotent }on $Y\subseteq X$ if $f(y_{N})=z$ for
each $y_{N}\in $ $Y^{N}$ such that $y_{i}=z$ for all $i\in N$.

A \textit{generalized committee }in $N$ is a set of coalitions $\mathcal{%
C\subseteq P}(N)$ such that $T\in \mathcal{C}$ if \ $T\subseteq N$ and $%
S\subseteq T$ for some $S\in \mathcal{C}$ ( a \textit{committee }in $N$
being a \textit{non-empty }generalized committee in $N$ which does \textit{%
not} include the \textit{empty} coalition)\footnote{%
Thus, a generalized committee is just an \textit{order filter} of the
partially ordered set $(\mathcal{P}(N),\subseteq )$ of coalitions of $N$.}.

A \textbf{generalized committee} \textbf{voting rule} is a function $f:\Pi
_{i\in N}Y_{i}\rightarrow X$ such that, for some fixed generalized committee 
$\mathcal{C\subseteq P}(N)$ and for all $y_{N}\in \Pi _{i\in N}Y_{i}$, $%
f(y_{N})=\vee _{S\in \mathcal{C}}(\wedge _{i\in S}x_{i})$.

A \textbf{generalized weak committee} \textbf{voting rule} is a function $%
f:\Pi _{i\in N}Y_{i}\rightarrow X$ such that, for some fixed generalized
committee $\mathcal{C\subseteq P}(N)$ and some fixed family $\left \{
z_{S}:z_{S}\in X\right \} _{S\in \mathcal{C}}$, and for all $y_{N}\in \Pi
_{i\in N}Y_{i}$, $f(y_{N})=\vee _{S\in \mathcal{C}}((\wedge _{i\in
S}x_{i})\wedge y_{S})$.

Two notable classes of strategy-proof voting rules are the \textit{%
projections }(or \textit{dictatorial rules) }$\pi _{i}:Y^{N}\rightarrow X$, $%
i\in N$ where for all $y_{N}\in Y^{N}$, $\pi _{i}(y_{N})=y_{i}$, and the 
\textit{constant }rules $f_{x}:$ $Y^{N}\rightarrow X$, $x\in X$ where for
all $y_{N}\in Y^{N}$, $f_{x}(y_{N})=x$. It is also easily checked that 
\textit{both dictatorial and constant rules are }$B_{\mathcal{X}}$-\textit{%
monotonic}.\footnote{%
Indeed, for all $x_{N}=(x_{j})_{j\in N}\in Y^{N},i\in N$ and $x_{i}^{\prime
}\in Y$: $f(x_{N})=x_{i}\in \lbrack x_{i},f(x_{i}^{\prime
},x_{N\smallsetminus \left \{ i\right \} })]$ if $f$ is the $i$-th
projection, and $f(x_{N})=f(x_{i}^{\prime },x_{N\smallsetminus \left \{
i\right \} })\in \lbrack x_{i},f(x_{i}^{\prime },x_{N\smallsetminus \left \{
i\right \} })]$ if $f$ is a constant function.}

The representation of $B_{\mathcal{X}}$\textit{-}monotonic voting rules as 
\textit{behaviour maps} of certain \textit{tree automata }acting on suitably 
\textit{labelled trees} will play a key role in the present paper. In that
connection, a few supplementary definitions are to be introduced here
concerning precisely certain tree automata and their behaviour (see Ad\'{a}%
mek and Trnkov\'{a} (1990) for a thorough treatment of tree automata).

A $\Sigma $-\textbf{tree automaton} is a general model of a mechanism that
can perform certain \textit{operations on its `internal' state space,} and
produce certain observable\textit{\ outputs} as a response to certain 
\textit{inputs }it is equipped to detect and act upon once it is suitably
prepared to do so, or \textit{initialized}. The \textit{operations} a tree
automaton is able to perform are algebraic or finitary (i.e. each of them
applies to some fixed finite number of arguments, its `arity'), and are
recorded together with their respective `arities' by the automaton's type,
denoted $\Sigma $. The \textit{inputs} of a tree automaton of a certain type 
$\Sigma $ are finite trees with some terminal nodes labelled by variables of
a certain set $I$ that have to be initialized, whereas all the other
terminal nodes are labelled by (symbols of) operations as recorded by type $%
\Sigma $ (such trees are also denoted here as finite $(\Sigma ,I)$-trees).
The \textit{initialization} of the automaton assigns one specific state to
every variable of $I$ and so makes it possible for the automaton to start
its action on any suitable tree-input. A tree-input dictates the admissible
(and mutually equivalent) sequences of operations to be performed by the
automaton as it inspects its nodes moving backward from the terminal nodes.
The final outcome of that sequence of operations or \textit{run }of the
initialized tree automaton produces as an end-result the state which is
computed performing the operation labelled by the initial node or \textit{%
root} of the tree as applied to the outcomes of its previous computations.
The output function then works as an `effector' that transforms the final
state thus obtained into an observable output. The \textit{behaviour }of a $%
\Sigma $-tree automaton denotes precisely the rule that transforms each $%
(\Sigma ,I)$-tree-input into a certain observable output through the process
just described (see Appendix 3 for a formal definition of $\Sigma $-tree
automata and all the relevant details).

Actually, we shall be concerned with \textit{latticial median -or }$l$%
\textit{-median- tree automata.} A (non-initial) \textbf{l-median tree
automaton}\textit{\ }$\mathcal{A}_{\mu }\mathcal{=}(X,\left\{ d_{s}\right\}
_{s\in \Sigma },Y,h)$ -also denoted as $\Sigma ^{\mu }$\textbf{-tree
automaton}\textit{- }is a (non-initial) $\Sigma $-tree automaton\textit{\ }%
with $\Sigma =\Sigma ^{\mu }$ comprising a unique ternary operation symbol $%
s_{\mu }$ denoting the median operation $\mu $ of a distributive lattice%
\footnote{%
Namely, a ternary operation on $X$ satisfying Birkhoff-Kiss axioms
m(i)-m(ii)-m(iii)-m(iv) (see Note 12 above).} $\mathcal{X}=(X,\leqslant )$
and a set of nullary operation symbols corresponding to some of the terminal
nodes of the labelled trees to be computed by the automaton, namely $\Sigma
^{\mu }=(\left\{ s_{\mu }\right\} \cup S_{0},\alpha )$ with $\alpha :\left\{
s_{\mu }\right\} \cup S_{0}\rightarrow \mathbb{Z}_{+}$ such that $\alpha
(s_{\mu })=3$, $\alpha (s)=0$ for each $s\in S_{0\text{ }}$and $d_{s_{\mu
}}=\mu $ . \ Given our present focus on a fixed population $N$ of agents of
size $n=|N|$ we may conveniently take $S_{0}$ such that $|S_{0}|=2+2^{n}$:
the elements of $S_{0}$ correspond to $0,1$ (standing, respectively, for the
bottom and top elements of the lattice) and to $2^{n}$ `phantom votes', one
for each coalition. We can also posit $Y=X$ and $h=id$ \ i.e. we take states
to be observable hence we can identify state and output spaces: it follows
that in the ensuing analysis we may safely identify, with a slight abuse of
language, the (non-initial) $\Sigma ^{\mu }$-tree automaton $\mathcal{A}%
_{\mu }$ and a $\Sigma ^{\mu }$-algebra on $X$. An \textit{initial} $\Sigma
^{\mu }$-tree automaton $\mathcal{A}_{\mu }^{I,\lambda }$ amounts to a $%
\Sigma ^{\mu }$-tree automaton as supplemented with an initialization $%
\lambda :I\rightarrow X$ i.e. an interpretation in $X$ of variables in $I$:
here, we take $|I|=n$, and the initialization $\lambda $ models a particular
ballot profile. Therefore, $\mathcal{A}_{\mu }^{I,\lambda }$ embodies an
interpretation in $X$ of all terminal nodes of \textit{any} finite labelled $%
(\Sigma ^{\mu },I)$-tree $T$ and is ready to compute an output of $T$ in $X$
-the \textit{behaviour }of $\mathcal{A}_{\mu }$ at $T$-as given by the value 
$\mathcal{A}_{\mu }^{I,\lambda }(T)$ of its run map at $T$. Thus, the
behaviour of median $\Sigma ^{\mu }$-tree automaton $\mathcal{A}_{\mu }$ at
any finite labelled $(\Sigma ^{\mu },I)$-tree $T$ \textit{is the outcome of
a nested sequence of medians} $\mu (........\mu (\mu (u,x_{i},z),x_{i},\mu
(u^{\prime },x_{i},z^{\prime })).......)$ starting with medians of
projections $x_{i}$ of $x_{N}$, $i=1,...,n$ and the $2^{n}$ elements of $%
S_{0}\subseteq X$ as dictated by $T$ in the following manner. Terminal nodes
of paths of maximum length $l$ come by construction in $2^{n-1}$ \textit{%
triples} that share an immediate predecessor labelled by $s_{\mu }$ i.e. the
symbol of the (latticial) median operation. The nodes of any such triple are
labelled by $x_{n}$ and two distinct elements of $S_{0}$. For any $k\leq l-1$%
, the terminal nodes of paths of length $l-k$ are labelled by $x_{n-k}$ (the
example below provides a very simple illustration with $n=3$).\medskip

\textbf{Example 10.} Let $\mathcal{X}=(X,\leqslant )$ be a distributive
lattice with bottom and top elements denoted by $\bot $ and $\top $\textit{,
respectively, and } $f:X^{3}\rightarrow X$ the\textbf{\ }voting rule for $%
(\left\{ 1,2,3\right\} ,X)$ defined as follows: for any $%
x=(x_{1},x_{2},x_{3})\in X^{3}$,

$f(x_{N})$ $=\mu (\mu (\mu (f(\bot ,\bot ,\bot ),x_{3},f(\bot ,\bot ,\top
)),x_{2},\mu (f(\bot ,\top ,\bot ),x_{3},f(\bot ,\top ,\top ))),$

$\ \ \ \ \ \ \ \ \ \ \ \ \ \ \ \ \ \ \ x_{1},\mu (\mu (f(\top ,\bot ,\bot
),x_{3},f(\top ,\bot ,\top )),x_{2},\mu (f(\top ,\top ,\bot ),x_{3},f(\top
,\top ,\top ))))$ .

Then, $f$ is l-median tree-automata representable: to see this, just
consider for any $x=(x_{1},x_{2},x_{3})\in X^{3}$ the corresponding labelled
tree\bigskip

Thus, we say that \textit{a voting rule is representable by a (finitary)
l-median tree automaton-}or\textit{\ \textbf{l-median tree-automata
representable (l-MTAR)} -} if it can be regarded as the behaviour of some
median $\Sigma $-tree automaton $\mathcal{A}_{\mu }$ as properly initialized
and acting on suitably labelled trees. That is made precise by the following:

\begin{definition}
(\textbf{l-median tree-automata representable (l-MTAR) voting rules}) Let $%
\mathcal{X}=(X,\leqslant )$ be a bounded distributive lattice. Then, a
voting rule $f:X^{N}\longrightarrow X$ \ is \textbf{l-median tree-automata
representable} \textbf{(l-MTAR)} if there exists a $\Sigma ^{\mu }$-tree
automaton\textit{\ }$\mathcal{A}_{\mu }$ such that for any $x_{N}\in X^{N}$
and any finite labelled $(\Sigma ^{\mu },I)$-tree $T$ there is a
corresponding \textit{initial} $\Sigma ^{\mu }$-tree automaton $\mathcal{A}%
_{\mu }^{I,\lambda }$ with $f(x_{N})=\mathcal{A}_{\mu }^{I,\lambda }(T)$.
\end{definition}

We are now ready to state the main result of this paper concerning the
characterization of strategy-proof voting rules on unimodal profiles. Our
characterization result relies on the following three lemmas.

The first lemma simply establishes the equivalence between $B_{\mathcal{X}}$-%
\textit{monotonicity} with respect to an arbitrary distributive lattice $%
\mathcal{X}$ and \textit{strategy-proofness }on the corresponding full
unimodal domain $U_{\mathcal{X}}^{N}$.

\begin{lemma}
Let $\mathcal{X}=(X,\leqslant )$ be a distributive lattice, and $%
f:X^{N}\rightarrow X$ a voting rule for $(N,X)$. Then, the following
statements are equivalent:

(i) $f$ is $B_{\mathcal{X}}$-\textit{monotonic;}

(ii) $f$ is strategy-proof on $U_{\mathcal{X}}^{N}$;

(iii) $f$ is strategy-proof on $S_{\mathcal{X}}^{N}$.
\end{lemma}

\textbf{Remark 3.}\ Lemma 1 above extends Lemma 1 of Danilov (1994)
(concerning linear orders in a tree that are unimodal with respect to
tree-betweenness). It also extends Proposition 3.2 of Nehring and Puppe
(2007 (a)) (concerning locally strictly unimodal domains in a \textit{finite}
distributive -actually, Boolean-lattice) since it holds for \textit{both}
the (full) unimodal and the (full) locally strictly unimodal domain in 
\textit{any distributive lattice (including infinite and non-boolean ones)}.
However, strictly speaking, Lemma 1 is not a generalization of Nehring and
Puppe's result since the latter concerns all social choice functions (not
just voting rules), and any `rich' locally strictly unimodal subdomain.

Observe that a restricted voting rule may be strategy-proof on its
restricted unimodal domain while being not monotonic (i.e. the implications
from (ii) or (iii) to (i) of the previous lemma do not hold in general for
restricted voting rules).

To see this, consider the following example, adapted from Barber\`{a}, Berga
and Moreno (2010), and slightly simplified: take $X=\left \{
a,b,c,d\right
\} $ with $a,b,c,d$ mutually distinct, $\Delta _{X}=\left \{
(x,x):x\in X\right \} $,

\begin{equation*}
\leqslant ^{\ast }=\left \{ (a,b),(a,c),(a,d),(b,c),(b,d),(d,c)\right \}
\cup \Delta _{X},
\end{equation*}%
i.e. $\mathcal{X}^{\ast }=(X,\leqslant ^{\ast })$ is the $4$-chain.

Then, posit $\ \ \ \ \ \ \ \ \ \ \ \ \succcurlyeq =(a\succ b\succ c\sim d)$

$\ \ \ \ \ \ \ \ \ \ \ \ \ \ \ \ \ \ \ \ \ \ \ \ \ \ \ \ \ \succcurlyeq
^{\prime }=(d\succ ^{\prime }b\succ ^{\prime }c\sim ^{\prime }a)$

\ \ \ \ \ \ \ \ \ \ \ \ \ \ \ \ \ \ \ \ \ \ \ \ \ \ \ \ \ \ $\succcurlyeq
^{\prime \prime }=(a\succ ^{\prime \prime }b\succ ^{\prime \prime }c\succ
^{\prime \prime }d)$

\ \ \ \ \ \ \ \ \ \ \ \ \ \ \ \ \ \ \ \ \ \ \ \ \ \ \ \ \ $\succcurlyeq
^{\prime \prime \prime }=(d\succ ^{\prime \prime \prime }b\succ ^{\prime
\prime \prime }c\succ ^{\prime \prime \prime }a)$

$D=\left \{ \succcurlyeq ,\succcurlyeq ^{\prime }\right \} $, $D^{\prime
}=\left \{ \succcurlyeq ^{\prime \prime },\succcurlyeq ^{\prime \prime
\prime }\right \} $, $Y=\left \{ a,d\right \} $ and define $f^{\prime
}:Y^{2}\times X^{N\smallsetminus \left \{ 1,2\right \} }\rightarrow X$ by
the following rule: for all $x_{N\smallsetminus \left \{ 1,2\right \} }\in
X^{N\smallsetminus \left \{ 1,2\right \} }$,%
\begin{eqnarray*}
f^{\prime }(a,a,x_{N\smallsetminus \left \{ 1,2\right \} }) &=&a\text{, \ \ }%
f^{\prime }(d,d,x_{N\smallsetminus \left \{ 1,2\right \} })=d\text{,} \\
\text{ }f^{\prime }(a,d,x_{N\smallsetminus \left \{ 1,2\right \} }) &=&b%
\text{, \ \ }f^{\prime }(d,a,x_{N\smallsetminus \left \{ 1,2\right \} })=c%
\text{.}
\end{eqnarray*}

First, observe that both $\succcurlyeq $ and $\succcurlyeq ^{\prime }$ are
in $U_{\mathcal{X}}^{N}$, i.e. are unimodal, while $\succcurlyeq ^{\prime
\prime }$and $\succcurlyeq ^{\prime \prime \prime }$ are locally strictly
unimodal: indeed, $top(\succcurlyeq )=top(\succcurlyeq ^{\prime \prime })=a$%
, $top(\succcurlyeq ^{\prime })=top(\succcurlyeq ^{\prime \prime \prime })=d$
and it is immediately seen that:%
\begin{equation*}
B_{\mathcal{X}}=\left\{ 
\begin{array}{c}
(a,b,c),(a,b,d),(a,d,c),(b,d,c), \\ 
(c,b,a),(d,b,a),(c,d,a),(c,d,b)%
\end{array}%
\right\} \cup \left\{ (x,y,z)\in X^{3}:x=y\text{ or }z=y\right\} \text{.}
\end{equation*}%
But then, since $\left\{ (b,c),(b,d),(d,c)\right\} \cup \Delta _{X}$ \ is a
subrelation of $\succcurlyeq $ and $\left\{ (b,c),(b,a),(d,a),(d,c)\right\}
\cup \Delta _{X}$ is a subrelation of $\succcurlyeq ^{\prime }$, it follows
that unimodality of $\succcurlyeq $ and $\succcurlyeq ^{\prime }$with
respect to $B_{\mathcal{X}}$ holds. Moreover, $f^{\prime }$ is by
construction strategy-proof on $D^{2}\times U_{\mathcal{X}}^{N\smallsetminus
\left\{ 1,2\right\} }$ (and on $(D^{\prime })^{2}\times S_{\mathcal{X}%
}^{N\smallsetminus \left\{ 1,2\right\} }$) : to check this, notice that $1$
and $2$ are the only non-dummy voters, and for all $x_{N\smallsetminus
\left\{ 1,2\right\} }\in X^{N\smallsetminus \left\{ 1,2\right\} }$,

$\ \ \ \ \ \ \ \ \ \ \ \ \ \ \ \ \ \ \ \ f^{\prime }(a,a,x_{N\smallsetminus
\left \{ 1,2\right \} })\succcurlyeq f^{\prime }(d,a,x_{N\smallsetminus
\left \{ 1,2\right \} })$, $f^{\prime }(a,d,x_{N\smallsetminus \left \{
1,2\right \} })\succcurlyeq f^{\prime }(d,d,x_{N\smallsetminus \left \{
1,2\right \} })$,

$\ \ \ \ \ \ \ \ \ \ \ \ \ \ \ \ \ \ \ f^{\prime }(a,a,x_{N\smallsetminus
\left \{ 1,2\right \} })\succcurlyeq f^{\prime }(a,d,x_{N\smallsetminus
\left \{ 1,2\right \} })$, $f^{\prime }(d,a,x_{N\smallsetminus \left \{
1,2\right \} })\succcurlyeq f^{\prime }(d,d,x_{N\smallsetminus \left \{
1,2\right \} })$,

and similarly

$\ \ \ \ \ \ \ \ \ \ \ \ \ \ \ \ \ f^{\prime }(d,a,x_{N\smallsetminus
\left
\{ 1,2\right \} })\succcurlyeq ^{^{\prime }}f^{\prime
}(a,a,x_{N\smallsetminus \left \{ 1,2\right \} })$,\ $f^{\prime
}(d,d,x_{N\smallsetminus \left \{ 1,2\right \} })\succcurlyeq ^{\prime
}f^{\prime }(a,d,x_{N\smallsetminus \left \{ 1,2\right \} })$,

$\ \ \ \ \ \ \ \ \ \ \ \ \ \ \ \ \ f^{\prime }(a,d,x_{N\smallsetminus
\left
\{ 1,2\right \} })\succcurlyeq ^{\prime }f^{\prime
}(a,a,x_{N\smallsetminus \left \{ 1,2\right \} })$, $f^{\prime
}(d,d,x_{N\smallsetminus \left \{ 1,2\right \} })\succcurlyeq ^{\prime
}f^{\prime }(d,a,x_{N\smallsetminus \left \{ 1,2\right \} })$,

whence strategy-proofness of $f^{\prime }$ on $D^{2}\times U_{\mathcal{X}%
}^{N\smallsetminus \left \{ 1,2\right \} }$follows (strategy proofness on $%
(D^{\prime })^{2}\times S_{\mathcal{X}}^{N\smallsetminus \left \{
1,2\right
\} }$ follows from the same argument by replacing $\succcurlyeq
^{\prime \prime }$and $\succcurlyeq ^{\prime \prime \prime }$for $%
\succcurlyeq $ and $\succcurlyeq ^{\prime }$, respectively).

However, observe that $f^{\prime }(d,a,x_{N\smallsetminus \left\{
1,2\right\} })=c\notin \lbrack d,a]=[d,f^{\prime }(a,a,x_{N\smallsetminus
\left\{ 1,2\right\} })]$ hence $f^{\prime }$ is \textit{not }$B_{\mathcal{X}%
} $-\textit{monotonic}.\bigskip

The next lemma ensures that in an arbitrary distributive lattice the median
operation as applied to voting rules does preserve $B_{\mathcal{X}}$%
-monotonicity.

\begin{lemma}
Let $\mathcal{X}=(X,\leqslant )$ be a distributive lattice, and$\
f:X^{N}\rightarrow X$, $g:X^{N}\rightarrow X$, $h:X^{N}\rightarrow X$ voting
rules that are $B_{\mathcal{X}}$-monotonic. Then $\mu
(f,g,h):X^{N}\rightarrow X$ (where $\mu (f,g,h)(x_{N})=\mu
(f(x_{N}),g(x_{N}),h(x_{N}))$ for all $x_{N}\in X^{N}$) is also $B_{\mathcal{%
X}}$-monotonic.
\end{lemma}

Finally, the next lemma - that only concerns \textit{bounded }distributive
lattices - provides a canonical median-based representation of all monotonic
voting rules hence - in view of Lemma 1 above - of all strategy-proof voting
rules on the corresponding full unimodal domain. That lemma relies on the
notion of a \textit{tree automaton} as defined above.

\begin{lemma}
Let $\mathcal{X}=(X,\leqslant )$ be a bounded distributive lattice and $%
f:X^{N}\rightarrow X$ a $B_{\mathcal{X}}$-\textit{monotonic voting rule}.
Then, $f$ is l-median tree-automata representable (l-MTAR).
\end{lemma}

The main implications of the foregoing lemmas are indeed summarized by the
following:

\begin{theorem}
Let $\mathcal{X}$ $=(X,\leqslant )$ be a bounded distributive lattice, $B_{%
\mathcal{X}}$ its latticial betweenness relation, and $f:X^{N}\rightarrow X$
a voting rule for $(N,X)$. Then, the following statements are equivalent:

(i) $f$ is $B_{\mathcal{X}}$-monotonic;

(ii) $f$ is strategy-proof on $U_{\mathcal{X}}^{N}$ \textit{;}

(iii) $f$ is strategy-proof on $S_{\mathcal{X}}^{N}$;

(iv) $f$ is l-MTAR;

(v) $f$ is a generalized weak committee voting rule\textbf{.}
\end{theorem}

\textbf{Remark 4.} Notice that Theorem 1 generalizes Moulin's
characterization of strategy-proof voting rules on (full) unimodal domains
in bounded chains to arbitrary bounded distributive lattices. Thus, it also
offers a direct extension to all bounded distributive lattices of Moulin's
original lattice-polynomial representation of strategy-proof voting rules to
be contrasted with the alternative characterization via families of
`left-coalition systems' on (full) locally strictly unimodal domains in
products of bounded chains due to Barber\`{a}, Gul and Stacchetti (1993),
which relies heavily on the product-structure of the underlying lattices. In
particular, Theorem 1 implies strategy-proofness of the simple majority
voting rule on unimodal domains (with an odd population of voters), since it
can be quite easily shown that the former is $B_{\mathcal{X}}$-monotonic
(see e.g. Monjardet (1990) for a formal definition and study of the simple
majority or extended median rule in a latticial framework). It follows that
in an arbitrary bounded distributive lattice there exist voting rules - such
as the simple majority rule - that jointly satisfy anonymity (i.e. symmetric
treatment of voters), neutrality (i.e. symmetric treatment of outcomes),
idempotence (i.e. faithful respect of unanimity of votes) and
strategy-proofness on the full unimodal domain.\bigskip

It can also be established, however, that strategy-proofness and coalitional
strategy-proofness of a voting rule are \textit{not }equivalent on unimodal
domains in bounded distributive lattices. This is made precise by the
following:

\begin{theorem}
Let $\mathcal{X}=(X,\leqslant )$ be a bounded distributive lattice. Then the
following holds:

$(i)$ if $|X|\geq 4$ then there exists a sublattice $\mathcal{Y=(}%
Y,\leqslant _{Y})$ of $\mathcal{X}$ (with $|Y|\geq 4$), subdomains $%
D\subseteq U_{\mathcal{X}}$ and $D^{\prime }\subseteq S_{\mathcal{X}}$, and
a restricted voting rule $f^{\prime }:Y^{2}\times X^{N\smallsetminus
\left
\{ 1,2\right \} }\rightarrow X$ that is strategy-proof on $%
D^{2}\times U_{\mathcal{X}}^{N\smallsetminus \left \{ 1,2\right \} }$%
\thinspace and on $(D^{\prime })^{2}\times S_{\mathcal{X}}^{N\smallsetminus
\left \{ 1,2\right
\} }$\thinspace but \textit{not} coalitionally
strategy-proof on $D^{2}\times U_{\mathcal{X}}^{N\smallsetminus \left \{
1,2\right \} }$\thinspace or on $(D^{\prime })^{2}\times S_{\mathcal{X}%
}^{N\smallsetminus \left \{ 1,2\right \} }$\thinspace ;

$(ii)$ if $|X|\geq 4$ and $\mathcal{X}$ is not a linear order then there
exists a sublattice $\mathcal{Y=(}Y,\leqslant _{Y})$ of $\mathcal{X}$ (with $%
|Y|\geq 4)$ and a voting rule $f^{\prime }:Y^{N}\rightarrow Y$ that is
strategy-proof \ on $U_{\mathcal{Y}}^{N}$ and on $S_{\mathcal{Y}}^{N}$ but 
\textit{not} coalitionally strategy-proof on $U_{\mathcal{Y}}^{N}$ or on $S_{%
\mathcal{Y}}^{N}$.\thinspace
\end{theorem}

Notice that if $f:X^{N}\rightarrow X$ is strategy-proof on $U_{\mathcal{X}%
}^{N}$ and $|X|\leq 3$ then $f$ is also coalitionally strategy-proof on $U_{%
\mathcal{X}}^{N}$: that implication follows from a straightforward
adaptation of the proof of Theorem 1 of Barber\`{a}, Berga and Moreno (2010)
to voting rules as combined with Proposition 1 of the same paper.

Moreover, as a further straightforward consequence of Theorem 2 (and of a
few previously known results), we have the following:

\begin{corollary}
Let $\mathcal{X}=(X,\leqslant )$ be a bounded distributive lattice. Then the
following statements are equivalent:

$(i)$ for each sublattice $\mathcal{Y}=(Y,\leqslant _{|Y})$ of $\mathcal{X}$
and each voting rule $f:Y^{N}\rightarrow Y$, $f$ is strategy-proof on $U_{%
\mathcal{Y}}^{N}$ (on on $S_{\mathcal{Y}}^{N}$, respectively) if and only if
it is also coalitionally strategy-proof on $U_{\mathcal{Y}}^{N}$ (on on $U_{%
\mathcal{Y}}^{N}$, respectively);

$(ii)$ $\mathcal{X}=(X,\leqslant )$ is a linear order.
\end{corollary}

Thus, we have here a remarkable characterization of \textit{bounded linear
orders} as \textit{the only bounded distributive lattices where equivalence
of individual and coalitional strategy-proofness of voting rules on full
unimodal domains holds}.

Indeed, the failure of equivalence between simple and coalitional
strategy-proofness pointed out by Theorem 2 is readily extended to an
impossibility result concerning availability of anonymous and idempotent
coalitionally strategy-proof voting rules for full unimodal domains(and
locally strictly unimodal domains) in a very general class of bounded
distributive lattices, even if (full) neutrality is dropped. That is made
precise by the following

\begin{theorem}
Let $\mathcal{X}=(X,\leqslant )$ be a bounded distributive lattice with at
least two distinct atoms $x,z\in X$ and $\mathcal{Y=(}Y,\leqslant _{Y})$ the
sublattice of $\mathcal{X}$ induced by the restriction of $\leqslant $ to $%
Y=\left \{ 0,x,z,x\vee z\right \} $. Then, there is no anonymous voting rule 
$f:X^{N}\rightarrow X$ which is locally sovereign and locally JI-neutral on $%
Y $, and coalitionally strategy-proof on $U_{\mathcal{X}}^{N}$ , or on $S_{%
\mathcal{X}}^{N}$.
\end{theorem}

Thus, in sharp contrast to what happens in chains, no anonymous
coalitionally strategy-proof voting rules are available on standard full
unimodal or locally strictly unimodal domains \textit{in bounded
distributive lattices with at least two atoms,} including Boolean $k$%
-hypercubes with $k>1$, even if an extended median-based aggregation rule is
well-defined, and (weak) efficiency or even (full) sovereignty are \textit{%
not} required at all.

\section{A simple example: single peakedness and strategy-proofness in the
Boolean square}

Consider a five-member committee $N=\left \{ 1,2,3,4,5\right \} $ facing a
decision problem concerning the formal requirements for candidates to fill a
certain top\ corporate position. The committee is to decide whether (1) Mild
-or Strict, i.e. more specific and demanding- formal qualifications and/or
(2) Medium -or High- seniority are to be required of candidates. The outcome
set is then $\left \{
(Mild,Medium),(Mild,High),(Strict,Medium),(Strict,High)\right \} $: denoting
both Mild and Medium by $0$ and both Strict and High by $1$ the outcome set
can be represented by the Boolean square $\mathbf{2}^{2}=(\left \{
0,1,x,y\right \} ,\leqslant )$ where $0=(0,0)$ denotes the bottom element, $%
1=(1,1)$ denotes the top element, while $x=(1,0)$ and $y=(0,1)$ are not
comparable. The latticial betweenness relation of $\mathbf{2}^{2}$ as
defined by the rule $[(a,c,b)\in B_{\mathbf{2}^{2}}$ iff $a\wedge b\leqslant
c\leqslant a\vee b]$ \ -where $\wedge $ and $\vee $ denote, respectively,
the meet and join of $\mathbf{2}^{2}$-\ is

$B_{\mathbf{2}^{2}}=\left \{ 
\begin{array}{c}
(0,x,1),(0,y,1),(0,0,1),(0,1,1),(1,x,0),(1,y,0),(1,0,0),(1,1,0), \\ 
(x,0,y),(x,1,y),(x,x,y),(x,y,y),(y,0,x),(y,1,x),(y,x,x),(y,y,x), \\ 
(0,0,x),(0,x,x),(x,0,0),(x,x,0),(0,0,y),(0,y,y),(y,0,0),(y,y,0), \\ 
(x,x,1),(x,1,1),(1,x,x),(1,1,x),(y,y,1),(y,1,1),(1,y,y),(1,1,y)%
\end{array}%
\right \} $ .

An unimodal total preorder for $\mathbf{2}^{2}$ is a total preorder $%
\succcurlyeq $ on $\left \{ 0,1,x,y\right \} $ with a unique maximum that
`respects' $B_{\mathbf{2}^{2}}$, namely such that for any $(a,c,b)\in B_{%
\mathbf{2}^{2}}$ either $c\succcurlyeq a$ or $c\succcurlyeq b$ (or both).
Then, if the complement of any element $a$ is denoted $a^{\prime }$, it is
easily checked that the unimodal total preorders on $\mathbf{2}^{2}$ are
precisely \textit{three} for each possible choice of the top outcome $a\in
\left \{ 0,x,y,1\right \} =\left \{ a,a^{\prime },b,b^{\prime }\right \} $
(hence \textit{twelve} altogether), namely

$\succcurlyeq _{1}\equiv a\succ _{1}b\succ _{1}b^{\prime }\sim _{1}a^{\prime
}$ ,\ \ \ \ \ \ $\succcurlyeq _{2}\equiv a\succ _{2}b^{\prime }\succ
_{2}b\sim _{2}a^{\prime }$\ \ ,\ \ \ \ \ \ $\succcurlyeq _{3}\equiv a\succ
_{3}b\sim _{3}b^{\prime }\sim _{3}a^{\prime }$\ .\ \ \ 

Indeed, the essential feature of an unimodal total preorder on $B_{\mathbf{2}%
^{2}}$ is simply the following: it must be the case that \textit{no
second-best is both a complement of the first-best and strictly better than
some other outcome.}

A voting rule $f$ is $B_{\mathbf{2}^{2}}$\textit{-monotonic} if -for any
agent $i$ and any profile $z_{-i}$ of the other agents' votes- $i$'s vote
for $u$ ensures an outcome $f(u,z_{-i})$ that lies between $u$ and $%
f(v,z_{-i})$ for any choice of $v$ in $\left\{ 0,1,x,y\right\} $.

The extended median or simple majority rule $\mu ^{\ast }:$\ $\left \{
0,1,x,y\right \} ^{5}\rightarrow \left \{ 0,1,x,y\right \} $ is defined as
follows: for any $(a_{1},a_{2},a_{3},a_{4},a_{5})\in \left \{
0,x,y,1\right
\} ^{5}$,

$\mu ^{\ast }(a_{1},a_{2},a_{3},a_{4},a_{5})=\vee _{T\subseteq N,|T|\geq
3}(\wedge _{i\in T}a_{i})$.

Since projections $f_{i}(a_{1},a_{2},a_{3},a_{4},a_{5})=a_{i}$ , $i=1,...,5$
are obviously $B_{\mathbf{2}^{2}}$-monotonic, and as shown below (see Lemma
2) the median preserves $B_{\mathbf{2}^{2}}$-monotonicity, it follows that
the median $\mu :\left \{ 0,1,x,y\right \} ^{3}\rightarrow \left \{
0,1,x,y\right \} $ as defined by the rule $\mu (a,b,c)=(a\wedge b)\vee
(b\wedge c)\vee (a\wedge c)$ is also $B_{\mathbf{2}^{2}}$-monotonic. (It can
also be easily checked that \textit{on the Boolean square} the median is
also efficient, since no outcome with zero votes can be selected by $\mu $:
that property, however, fails in higher dimensional Boolean hypercubes).

Now, it turns out that both the join and the meet of any $(a,b)\in \left \{
0,x,y,1\right \} ^{2}$ (and therefore the join and the meet of any finite
subset of $\left \{ 0,x,y,1\right \} $) are representable as the median
(iterated median, respectively) of two projections and one constant, namely

$a\vee b=(a\wedge b)\vee (b\wedge 1)\vee (a\wedge 1)=\mu (a,b,1)$ and

$a\wedge b=(a\wedge b)\vee (b\wedge 0)\vee (a\wedge 0)=\mu (a,b,0)$.

Since constants (when regarded as constant functions) are obviously $B_{%
\mathbf{2}^{2}}$-monotonic, it also follows (from the $B_{\mathbf{2}^{2}}$%
-monotonicity preservation property of the median as mentioned above) that
joins and meets of any finite subset of $\left \{ 0,x,y,1\right \} $ may be
reduced to iterated medians of projections and constants, and are therefore
also $B_{\mathbf{2}^{2}}$-monotonic. The same argument applies in particular
to the extended median or simple majority rule $\mu ^{\ast }$ to conclude
that $\mu ^{\ast }$ is $B_{\mathbf{2}^{2}}$-monotonic (with $n=5$ voters a
direct check is of course still manageable, but the computation starts to
become quite long and tedious).

Thus, since Theorem 1 implies that the strategy-proof voting rules for $%
(N,\left \{ 0,x,y,1\right \} )$ on the full unimodal domain of total
preorders\ in the Boolean square are precisely the $B_{\mathbf{2}^{2}}$%
-monotonic\ functions on $\left \{ 0,x,y,1\right \} ^{N}$, it follows that $%
\mu ^{\ast }$ itself is in fact strategy-proof on that unimodal domain
(along with all the $B_{\mathbf{2}^{2}}$-monotonic\ functions on $\left \{
0,x,y,1\right \} ^{5}$). \ 

Next, consider the following profile of total preorders:

$\succcurlyeq _{1}=\succcurlyeq _{2}\equiv (x\succ 1\succ 0\sim y)$, \ \ $%
\succcurlyeq _{3}=\succcurlyeq _{4}\equiv (y\succ 1\succ 0\sim x)$, \ $%
\succcurlyeq _{5}\equiv (0\succ x\succ y\sim 1)$.

That profile is obviously unimodal with respect to $B_{\mathbf{2}^{2}}$ (see
the definition above) since no second-best outcome is a complement of its
first-best, and it is easily checked that

$\mu ^{\ast }(x,x,y,y,0)=0,$ whereas $\mu ^{\ast }(1,1,1,1,0)=1$.

It follows that coalition $\left \{ 1,2,3,4\right \} $ can successfully
manipulate $\mu ^{\ast }$ at that preference profile, namely the simple
majority rule is \textit{not coalitionally strategy-proof }on the full
unimodal domain with respect to the latticial betweenness relation $B_{%
\mathbf{2}^{2}}$ (Theorem 2 shows that such a situation always occurs
whenever the underlying bounded distributive lattice is not a chain). And
all of the above can be generalized to an impossibility result: no anonymous
voting rule enjoying a modicum of neutrality/sovereignty is coalitionally
strategy-proof on unimodal domains in bounded distributive lattices that
-like finite Boolean lattices (or hypercubes) $\mathbf{2}^{m}$, $m\geq 2$,
and \textit{unlike }chains- have at least two distinct atoms i.e. two
elements that cover the bottom element (this is the content of Theorem 3).

To the the best of the authors' knowledge, none of those results on the
Boolean \textit{square} is available in the previous literature on
strategy-proofness and single peakedness, and the same holds for
counterparts of them under other notions of single peakedness. Partial
results implying or suggesting some counterparts of Theorems 2 and 3 for
finite Boolean lattices $\mathbf{2}^{m}$, $m\geq 3$ under \textit{alternative%
} notions of single peakedness are indeed available (including results on 
\textit{separable preferences}, that are usually \textit{not} presented as
an instance of a single peaked domain but \textit{can be}, as shown below):
see e.g. Nehring and Puppe (2007(a),(b)), and Barber\'{a}, Sonnenschein and
Zhou (1991). It should be stressed again, however, that such results -while
interesting and valuable by themselves- are independent and at least in one
respect narrower than those presented in the present work. That is so
because the former not only ignore at all the unimodal case, but even in the
locally strict unimodal case fail to cover -as opposed to Theorems 2 and 3
below- infinite lattices and the Boolean square (i.e. the finite Boolean
case with $m=2$). Furthermore, it should be stressed that locally strict
unimodality is somewhat at odds with a latticial outcome set. That is so
because it relies on a notion of proximity-based betweenness-consistency
that \textit{cannot be backed by any metric} \textit{consistent with
standard latticial betweenness relations without allowing for multiple local
peaks.}

In order to clarify those statements, it is worth reviewing here the sort of
betweenness relations underlying such alternative notions of single
peakedness.

The most widely used alternative version of single peakedness encountered in
the extant literature requires that (i) there exist a unique maximum or best
outcome, and (ii) any outcome $x$ that lies between the best outcome and
another outcome $y$ distinct from $x$ itself should also be \textit{strictly
better} than $y$ (see e.g. Barber\`{a}, Gul and Stacchetti (1993), and
Nehring and Puppe (2007 (a,b)). It is easily checked that under such notion
of single peakedness (labeled \textit{`general single peakedness' }in
Nehring and Puppe (2007 (b)) and \textit{`locally strict unimodality'} in
the present paper) as applied to (the relevant part of) latticial
betweenness, single peaked total preorders on $\mathbf{2}^{2}$ are -again- 
\textit{three} for each possible choice of the top outcome $a\in \left \{
0,x,y,1\right \} =\left \{ a,a^{\prime },b,b^{\prime }\right \} $ (hence 
\textit{twelve} altogether), namely

$\succcurlyeq _{1}^{\prime }\equiv a\succ _{1}^{\prime }b\succ _{1}b^{\prime
}\succ _{1}^{\prime }a^{\prime }$ ,\ \ \ \ \ \ $\succcurlyeq _{2}^{\prime
}\equiv a\succ _{2}^{\prime }b^{\prime }\succ _{2}^{\prime }b\succ
_{2}^{\prime }a^{\prime }$\ \ ,\ \ \ \ \ \ $\succcurlyeq _{3}^{\prime
}\equiv a\succ _{3}^{\prime }b\sim _{3}^{\prime }b^{\prime }\succ
_{3}^{\prime }a^{\prime }$\ .\ 

Thus, as it is immediately checked, unimodal and locally strict unimodal
total preorders comprise two \textit{disjoint} sets.

Indeed, under locally strict unimodality, it is still possible to claim that
every single voter's preferences are `consistent' with the same betweenness
relation, namely the latticial betweenness $B_{\mathbf{2}^{2}}$. However,
the implied `consistency' is formulated in such a way that \textit{only
certain parts of }$B_{\mathbf{2}^{2}}$\textit{\ play an active role in
shaping preferences}, and \textit{distinct parts of it play such an active
role for agents having distinct top outcomes.} In particular, and most
remarkably, for each locally strictly unimodal total preorder $\succcurlyeq $
on $\mathbf{2}^{2}$ with top outcome $a$ both $b\succ a^{\prime }$ and $%
b^{\prime }\succ a^{\prime }$ hold, while of course $(b,a^{\prime
},b^{\prime })\in B_{\mathbf{2}^{2}}$: therefore, when applied to $B_{%
\mathbf{2}^{2}}$ in its entirety, locally strictly unimodal total preorders
actually admit \textit{two }local peaks. That fact strongly suggests that 
\textit{if} locally strictly unimodal total preorders are to be regarded as 
\textit{single peaked }with respect to \textit{some }betweenness relation on 
$\mathbf{2}^{2}$, then claiming that role for the \textit{entire }latticial
betweenness $B_{\mathbf{2}^{2}}$, while being of course a legitimate
stipulation is \textit{arguably somewhat far-fetched}. A far more natural
and appropriate choice for that role would be apparently its proper
subrelation $B_{\mathbf{2}^{2}}^{a}=B_{\mathbf{2}^{2}}\smallsetminus \left\{
(b,a,b^{\prime }),(b^{\prime },a,b)\right\} $. Observe, however, that such
an approach would result in a `non-classic' notion of single peakedness that
makes reference to \textit{several} preference-dependent hence, generally
speaking, \textit{agent-dependent betweenness relations (one for each
possible best outcome)}.

To be sure, there is still another possibility to anchor all locally
strictly unimodal total preorders to\textit{\ }a \textit{common betweenness
relation}: that would entail choosing \textit{another }proper subrelation of 
$B_{\mathbf{2}^{2}}$, namely $B_{\mathbf{2}^{2}}^{\ast }=B_{\mathbf{2}%
^{2}}\smallsetminus \left \{ (x,1,y),(y,1,x),(x,0,y),(y,0,x)\right \} $ as
the relevant betweenness relation. Notice that $B_{\mathbf{2}^{2}}^{\ast }$
is in fact the natural betweenness relation of $\mathbf{2}^{2}$ when
regarded not as a (Boolean) lattice, but \textit{just as a partially ordered
set}: thus, outcome $b$ is declared to lie between $a$ and $c$ if and only
if either $a\leqslant b\leqslant c$ or $c\leqslant b\leqslant a$ hold (the
hallmark of order betweenness is that \textit{no third outcome lies between
two incomparable elements}). That move would enlarge the set of locally
strictly unimodal total preorders, collapsing locally strict unimodality to
uniqueness of the best outcome \textit{whenever the best outcome is either }$%
x$ or $y$: in that case, we would end up with a notion of single peakedness
for total preorders that relies on a unique shared betweenness relation but
is itself \textit{essentially preference-dependent} anyway. In any case,
that choice of the relevant betweenness would be at variance with a full
fledged treatment of outcome set $\mathbf{2}^{2}$ as a (distributive) 
\textit{lattice. }It would result in a considerable relaxation of single
peakednees restrictions, and a strengthening of the relevant notion of
monotonicity, to the effect of rendering the median function \textit{not}
monotonic.\footnote{%
To see this, consider for instance $(x,y,1)$ and $(y,y,1)$. Clearly, $\mu
(x,y,1)=1$ , $\mu (y,y,1)=y$ and \textit{not }$B_{\mathbf{2}^{2}}^{\ast
}(x,1,y)$ hence $\mu $ is not $B_{\mathbf{2}^{2}}^{\ast }$-monotonic, and no 
$B_{\mathbf{2}^{2}}^{\ast }$-counterpart of Theorem 1 below applies to it.}

\textit{Separable preferences }on $\mathbf{2}^{2}$ (a notion due to Barber%
\'{a}, Sonnenschein and Zhou (1991) who define it for arbitrary finite
Boolean lattices) are best introduced by regarding $\mathbf{2}^{2}$ as the
power set of a two-item set $\left \{ x,y\right \} $ with $x$ and $y$
denoting singletons $\left \{ x\right \} $ and $\left \{ y\right \} $, and
the empty set $\varnothing $ and $\left \{ x,y\right \} $ itself \ standing
for $0$ and $1$, respectively. Any item $a\in \left \{ x,y\right \} $ is
either `good' or `bad': it is \textit{good} if $\left \{ a\right \} \succ
\varnothing $ and \textit{bad }if $\varnothing \succ \left \{ a\right \} $.
The set of all good items of a total preorder $\succcurlyeq $ on $\mathbf{2}%
^{2}$ is denoted by $G(\succcurlyeq )$. A total preorder $\succcurlyeq $ on $%
\mathbf{2}^{2}$ is \textit{separable } if for any $A\subseteq \left \{
x,y\right \} $ and $a\in \left \{ x,y\right \} \smallsetminus A$, $A\cup
\left \{ a\right \} \succ A$ if and only if $a\in G(\succcurlyeq )$.
Clearly, for any separable total preorder $\succcurlyeq $ on $\mathbf{2}^{2}$%
, $G(\succcurlyeq )$ is the unique maximum of $\succcurlyeq $. Resuming now
for the sake of comparisons the standard notation used in the former
discussion, the separable total preorders are three for each possible choice
of the best outcome (that is, recall, the set of all good items), namely

$\succcurlyeq _{1}^{\prime \prime }\equiv a\succ _{1}^{\prime \prime }b\succ
_{1}^{\prime \prime }b^{\prime }\succ _{1}^{\prime \prime }a^{\prime }$ ,\ \
\ \ \ \ $\succcurlyeq _{2}^{\prime \prime }\equiv a\succ _{2}^{\prime \prime
}b^{\prime }\succ _{2}^{\prime \prime }b\succ _{2}^{\prime \prime }a^{\prime
}$\ \ ,\ \ \ \ \ \ $\succcurlyeq _{3}^{\prime \prime }\equiv a\succ
_{3}^{\prime \prime }b\sim _{3}^{\prime \prime }b^{\prime }\succ
_{3}^{\prime \prime }a^{\prime }$

where $a$ denotes the set of all good items, $a^{\prime }$ is the complement
of $a$, and $b^{\prime }$ is the complement of $b$.

Notice that on $\mathbf{2}^{2}$ separable preferences are isomorphic to
locally strictly unimodal preferences.\footnote{%
The argument can be extended to arbitrary Boolean hypercubes $\mathbf{2}^{m}$%
, $m\geq 2$.} Thus, separable preferences are just locally strictly unimodal
preferences on finite Boolean lattices in disguise. It follows that the same
observations made on the latter also apply to separable preferences:
precisely as locally strictly unimodal preferences, separable preferences
can be regarded as single peaked either with respect to multiple,
agent-dependent betweenness relations or with respect to a common
betweenness relation. In both cases, however, \textit{the betweenness
relations involved and playing an active role depend on preferences, are
distinct from latticial betweenness }$B_{\mathbf{2}^{2}}$ and, arguably, 
\textit{do disregard in relevant ways the latticial structure of the outcome
set}.

\section{Related literature and concluding remarks}

The main results of the present paper may be summarized as follows:

$(i)$ Theorem 1 provides a characterization in terms of iterated medians of
projections and constants of the class of strategy-proof voting rules on
(full) unimodal domains and locally strictly unimodal domains \textit{in all
bounded distributive lattices}: thus, combining a version of the original
Moulin's lattice-polynomial representation with a suitable generalization of
ideas and techniques proposed by Danilov (1994) through an explicit reliance
on tree automata, it extends in significant ways both Moulin (1980) and
Danilov (1994) (which only concern \textit{unimodal domains} in \textit{%
bounded chains} and in \textit{bounded trees}, respectively), and Barber\`{a}%
, Gul and Stacchetti (1993) and Nehring and Puppe (2007 (a),(b)) (which only
concern \textit{locally strictly unimodal domains }in \textit{finite
products of bounded chains} and in \textit{finite }distributive lattices,
respectively).

$(ii)$ Theorem 2 establishes that \textit{equivalence between (individual)
strategy-proofness and coalitional strategy-proofness on both full unimodal
and full locally strictly unimodal domains} holds precisely in bounded
linear orders, and \textit{fails in bounded distributive lattices that are
not linear orders}: it complements the opposite results obtained by Moulin
(1980) and Danilov (1994) for full unimodal domains in bounded chains and
trees and by Barber\`{a}, Berga and Moreno (2010) for locally strictly
unimodal domains in bounded chains, and extends previous results obtained by
Barber\`{a}, Sonnenschein and Zhou (1991) and Nehring and Puppe (2007
(a),(b)) for locally strictly unimodal domains in certain \textit{finite}
distributive lattices.

$(iii)$ Theorem 3 establishes -for bounded distributive lattices with at
least two atoms- the impossibility of anonymous coalitional strategy-proof
voting rules with even a minimal amount of local sovereignty and local
neutrality on full unimodal domains: it also extends to a large subclass of
non-sovereign voting rules for full unimodal and locally strictly unimodal
domains in a much larger class of bounded distributive lattices some
previous results mainly due to Barber\`{a}, Gul and Stacchetti (1993) and
Nehring and Puppe (2007 (a), (b)) concerning voting rules for locally
strictly unimodal domains in certain \textit{finite} or \textit{product}
distributive lattices (indeed, a distributive lattice with two or more atoms
need not be a product lattice, or finite).

In order to properly appreciate the significance of the foregoing result a
few key contributions from the early literature on related issues are to be
discussed in some detail.

The seminal paper by Moulin (see Moulin (1980)) provides an explicit
characterization in terms of `extended medians' of the class of all
strategy-proof voting rules on the domain of \textit{all} profiles of total
preorders that are unimodal with respect to a fixed \textit{bounded linear
order\footnote{%
To be sure, Moulin proves the characterization result mentioned above for a 
\textit{restricted }unimodal domain where voters are not allowed to regard
the maximum or the minimum of the chain as their unique optimum. But
Moulin's proof can be adapted to the full unimodal domain.}}. Furthermore,
Moulin (1980) establishes the equivalence of strategy-proofness and
coalitional strategy-proofness for \textit{all }voting rules on such full
unimodal domains. Clearly, Moulin's result does not apply to the Boolean
square. Moreover, its proof cannot be extended to the latter.

In fact, Moulin's proof relies heavily on the following property of medians
in bounded linear orders that \textit{does not hold for medians in general
bounded distributive lattices}: given an odd population of $n=2k+1$ voters,
for any $(x_{i})_{i=1,...,n}\in X^{N}$ the (extended) median $\mu ^{\ast
}(x_{1},...,x_{n})$ i.e. the (iterated) median $\mu (x_{2k},\mu
(x_{2(k-1)},\mu (..(\mu (x_{1},x_{2},x_{3}))..),x_{2k-1}),x_{2k+1})$ is such
that:%
\begin{equation}
\min (\left \vert \left \{ i\in N:x_{i}\leq \mu ^{\ast
}(x_{1},...,x_{n})\right \} \right \vert ,\left \vert \left \{ i\in N:\mu
^{\ast }(x_{1},...,x_{n})\leq x_{i}\right \} \right \vert )\geq k+1\text{.}
\label{ste}
\end{equation}

However, take $n=3$ (hence $k+1=2$) and consider the Boolean square $\mathbf{%
2}^{2}$.

Clearly $\mu ^{\ast }(1,x,y)=\mu (1,x,y)=1$, hence at $%
(x_{1},x_{2},x_{3})=(1,x,y)$, $\left \vert \left \{ i\in N:x_{i}\leq \mu
^{\ast }(1,x,y)\right \} \right \vert =3$, but $\left \vert \left \{ i\in
N:\mu ^{\ast }(1,x,y)\leq x_{i}\right \} \right \vert =1$, and \ref{ste}
fails.

In a similar vein, Danilov (1994) provides a characterization in terms of
(iterated) medians of the class of strategy-proof voting rules on the domain
of all unimodal \textit{linear orders} (i.e. \textit{antisymmetric} total
preorders) when $X$ is the vertex set of an undirected (bounded) \textit{%
tree }(see also Danilov and Sotskov (2002) for further discussion of this
topic, and Demange (1982) for an early study of majority-like voting rules
on domains of unimodal linear orders in undirected trees).\footnote{%
Danilov's result can also be reformulated in terms of median tree-automata
representable voting rules by suitably redefining the median operation in
order to reflect the characteristic properties of medians of trees (as
opposed to medians of bounded distributive lattices). This can be done
replacing axiom $m(i)$ of Birkhoff and Kiss (1947) (see note 12 above) with
\par
$m^{\prime }(i)$: for all $a,b,c,d$,
\par
either $m(m(d,a,b),c,d)=m(d,a,c)$%
\par
or $m(m(d,a,b),c,d)=m(d,b,c)$%
\par
(see Sholander (1952)).}\textit{\ }Moreover, Danilov (1994) also shows that
strategy-proofness and coalitional strategy-proofness of voting rules on 
\textit{unimodal profiles of linear preference orders in undirected bounded
trees} are \textit{equivalent} properties. But in fact, it can be shown that
Danilov's proofs can be readily extended to the wider full domain of \textit{%
unimodal total preference preorders }(arguing along the lines of the first
part of the proof of Lemma 1 above), and to the case of an underlying 
\textit{bounded linearly ordered set }of alternatives.

However, the key step of Danilov's proof relies on the following property
shared by \textit{intervals }of linear orders and of undirected trees,
namely:

\begin{equation}
\text{for all }x,y,v,z\in X\text{ such that }x\neq y\text{, if }x\in \lbrack
y,v]\text{ and }y\in \lbrack x,z]\text{ then }x\in \lbrack v,z]\text{.}
\label{ste1}
\end{equation}%
Notice however that \ref{ste1} does \textit{not }hold for (latticial)
intervals of arbitrary bounded distributive lattices. To see this, consider
again precisely the Boolean square $\mathbf{2}^{2}$ and notice that e.g. $%
b\in \lbrack a,d]$, $a\in \lbrack b,c]$ but $b\notin \lbrack c,d]$.\footnote{%
See e.g. Sholander (1952, 1954(a)), Bandelt and Hedl\'{\i}kov\'{a} (1983)
for a thorough study of intervals in general median algebras, and Isbell
(1980) for an even more general approach that also considers intervals in a
larger class of ternary algebras.}

Moulin's fundamental characterization result also inspired many authors to
explore related strategy-proofness issues on \textit{other} single peaked
domains (including locally strictly unimodal domains). Those alternative
notions of single peakedness typically rely on binary proximity relations.
Such proximity relations are usually jointly induced together \textit{with}
the betweenness relations \textit{by }an underlying agent-invariant \textit{%
metric }structure hence are agent-invariant and total (see e.g. Border and
Jordan (1983), Barber\`{a}, Gul and Stacchetti (1993), Chichilnisky and Heal
(1997), Ching (1997), Barber\`{a}, Mass\`{o} and Neme (1997), Peremans,
Peters, van der Stel and Storcken (1997), Schummer and Vohra (2002), Nehring
and Puppe (2007(a), 2007(b)), Bordes, Laffond and Le Breton (2012),
Chatterji, Sanver and Sen (2013)). Those works include several
characterization results concerning some large subclass of strategy-proof
voting mechanisms on suitably defined single-peaked domains (such as
sovereign i.e. surjective voting rules or social choice functions).
Moreover, some of those results provide valuable information on
strategy-proofness properties of voting rules for certain proximity-based
single peaked domains in certain distributive lattices and provide partial
counterparts to our results on coalitional manipulability of median-based
rules.

Thus, the existence of \textit{sovereign} strategy-proof voting rules for
Euclidean single peaked domains that are not coalitionally strategy-proof\
is well-established for proximity-based single peaked profiles in Euclidean $%
m$-dimensional spaces with $m\geq 2$ (see e.g. Border and Jordan (1983),
Peters, van der Stel and Storcken (1992), Peremans, Peters, van der Stel and
Storcken (1997), Bordes, Laffond and Le Breton (2012)). Furthermore,
Euclidean spaces are (unbounded) Riesz spaces i.e. are endowed with a
natural (unbounded) distributive latticial structure. Notice however that
the betweenness relation induced by the Euclidean metric is distinct from
the latticial betweenness relation and in fact no well-behaved median
operation is available in an $m$-dimensional Euclidean space with $m\geq 2$.

In their influential contribution, Barber\`{a}, Gul and Stacchetti (1993)
identify single peakedness and locally strict unimodality and offer an
alternative characterization of strategy-proof voting rules on (full)
locally strictly unimodal domains in \textit{(finite) products of bounded
chains endowed with the }$L_{1}$\textit{-metric} \footnote{%
Namely, $d^{L_{1}}(x,y)=\Sigma _{i}|x_{i}-y_{i}|$ for all $x,y\in X$. The $%
L_{1}$-metric is consistent with latticial betweenness in the sense
explained in Remark 1 above. Barber\`{a}, Gul and Stacchetti (1993) also
provides a characterization of strategy-proof social choice rules with range
given by a product of sub-chains.}: their characterization relies on \textit{%
generalized median voter schemes} i.e. on representations of voting rules
via families of outcome/dimension-specific generalized committees denoting
winning coalitions. While that work mainly focusses on the finite case, it
can be readily extended to finite products of arbitrary bounded chains.
However, it relies heavily on the product structure of the underlying
lattices.

Building upon some remarkable earlier contributions including Barber\`{a},
Gul and Stacchetti (1993) and Barber\`{a}, Mass\`{o} and Neme (1997), and
also focusing on \textit{locally strict} unimodality (under the label `%
\textit{generalized single peakedness')}, Nehring and Puppe (2007(a)) offer
a comprehensive study and an `issue-by-issue voting-by-committees'-based
characterization of sovereign (i.e. surjective) strategy-proof voting rules
on rich domains of so-called\ locally strictly unimodal \ \textit{linear} 
\textit{orders} in certain \textit{finite} \textit{median interval spaces}%
\footnote{%
A median interval space amounts to a set with a ternary betweenness relation
such that for any triple of elements there exists precisely one element
which lies between each pair in the triple.} as induced by suitably defined
`property spaces'. In particular, Nehring and Puppe (2007 (a), (b)) provide
valuable results on locally strict unimodality and strategy-proofness in 
\textit{finite} median spaces, and (finite) distributive lattices \textit{%
are }a prominent instance of (finite) median spaces. However, due to their
choice of \textit{linear }preference domains as combined with their
(locally) \textit{strict }notion of unimodality, it turns out that their
results are in fact irrelevant for the case of \textit{unimodality} in
finite distributive lattices other than finite chains, as explained in some
detail in Section 2 above. Furthermore, Nehring and Puppe (2007 (b)) prove
that the only efficient and strategy-proof voting rules on `rich' domains of
locally strictly unimodal profiles of linear orders in finite Boolean $m$%
-hypercubes with $m\geq 3$ are weakly dictatorial. Notice, however, that
efficient voting rules are in particular \textit{sovereign} hence that
result is not in any case an impossibility result for non-sovereign
anonymous and coalitionally strategy-proof voting rules. It should also be
stressed that the main result in Nehring, Puppe (2007 (b)) does entail
equivalence-failure for simple and coalitional strategy-proofness in Boolean 
$k$-hypercubes for $k\geq 3$ but it refers to a domain of locally strictly
unimodal linear orders that is distinct -and in fact \textit{disjoint }in
any Boolean hypercube-\textit{\ }from the domain of all unimodal total
preorders which is the focus of the present work. Moreover, even on the
locally strictly unimodal\textit{\ }domain the foregoing result does not
apply to the Boolean \textit{square}, while our Theorem 2 covers the full
unimodal domain in both infinite bounded distributive lattices and arbitrary
Boolean hypercubes, including of course the Boolean square $\mathbf{2}^{2}$.

Another recent paper (Chatterji, Sanver and Sen (2013)) provides a
characterization of those \textit{`strongly-path-connected'} domains of 
\textit{linear orders} that ensure existence of anonymous and idempotent
strategy-proof social choice functions for a voter population of \textit{even%
} size: such characterization relies on a new, generalized notion of
single-peakedness for linear orders denoted as `semi-single-peakedness'
(requiring essentially that the outcome set can be endowed with a tree
structure such that locally strict unimodality as defined above holds within
a certain threshold-distance from the top outcome). However, the domain of
(all) unimodal \textit{linear} orders on the Boolean square is -as observed
above- empty hence it is trivially \textit{not strongly path-connected}.
Thus, the unimodal domain is definitely beyond the scope of the
Chatterji-Sanver-Sen characterization.

Barber\`{a}, Berga and Moreno (2010) addresses the general issue of
equivalence between simple and coalitional strategy-proofness and consider 
\textit{locally strictly unimodal domains} of total preorders.\footnote{%
It should be noticed, however, that locally strict unimodality in a bounded
linear order reduces to unimodality when total preorders are in fact
antisymmetric i.e. linear orders: details are available from the authors
upon request.} They establish that a property they newly introduce and label
`\textit{Sequential Inclusion}' provides a general sufficient condition
ensuring equivalence of individual and coalitional strategy-proofness, and
show that \textit{locally} \textit{strictly unimodal domains of total
preorders} as defined on a \textit{linear order }$(X,\leqslant )$ do satisfy
it. Specifically, for each preference profile $(\succcurlyeq _{i})_{i\in N}$ 
\textit{Sequential Inclusion} relies on a family of binary relations $%
\succcurlyeq (S((\succcurlyeq _{i})_{i\in N},y,z))$ as parameterized by
ordered pairs $(y,z)$ of outcomes and defined on $S((\succcurlyeq
_{i})_{i\in N},y,z)$, the set of voters that strictly prefer $y$ to $z$ at $%
(\succcurlyeq _{i})_{i\in N}$: in particular, voter pair $(i,j)$ is in $%
\succcurlyeq (S((\succcurlyeq _{i})_{i\in N},y,z))$ if and only if $i$ and $%
j $ are in $S((\succcurlyeq _{i})_{i\in N},y,z)$ and $\left\{ x\in
X:z\succcurlyeq _{i}x\right\} \subseteq \left\{ x\in X:z\succ _{j}x\right\} $%
. Of course any such $\succcurlyeq (S((\succcurlyeq _{i})_{i\in N},y,z))$ is
reflexive: Sequential Inclusion requires that all of them be also \textit{%
connected and acyclic. Indirect Sequential Inclusion} is satisfied by a
profile $(\succcurlyeq _{i})_{i\in N}$ if \textit{either} $(\succcurlyeq
_{i})_{i\in N}$ itself satisfy Sequential Inclusion \textit{or} for each
pair $(y,z)$ of outcomes there exists a profile $(\succcurlyeq _{i}^{\prime
}:$ $i\in S((\succcurlyeq _{i})_{i\in N},y,z))$ such that: (i) $y\succ
_{i}^{\prime }z$ for each $i\in S((\succcurlyeq _{i})_{i\in N},y,z)$, (ii) $%
z\succ _{i}^{\prime }x$ for each $i\in $ $S((\succcurlyeq _{i})_{i\in
N},y,z) $ and each outcome $x\neq z$ such that $z\succcurlyeq _{i}x$, and
(iii) $\succcurlyeq (S((\succcurlyeq _{i}^{\prime }:i\in S((\succcurlyeq
_{i})_{i\in N},y,z)))$ is connected and acyclic. A preference domain is then
said to satisfy \textit{Sequential Inclusion (Indirect Sequential Inclusion) 
}if each preference profile in that domain does satisfy it.\footnote{%
Moreover, it turns out that in our full unimodal setting Indirect Sequential
Inclusion is a generalization of a similar `richness' condition singled out
by Le Breton and Zaporozhets (2009).}

Thus, the Barber\`{a}-Berga-Moreno result mentioned above extends to locally
strictly unimodal domains Moulin's equivalence between individual and
coalitional strategy-proofness on unimodal domains in bounded chains (see
also Danilov and Sotskov (2002) and Le Breton and Zaporozhets (2009) in that
connection). Notice, incidentally, that the Barber\`{a}-Berga-Moreno
argument for such an equivalence result \textit{cannot be extended to
domains of unimodal total preorders even in bounded linear orders}.\footnote{%
Indeed, take a four-element \textit{linear order }$(\left\{ x,y,w,z\right\}
,\leqslant )$ such that $x<y<z<w$ , consider total preorders $\succcurlyeq
_{1},\succcurlyeq _{2}$ on $X$ such that $x\succ _{1}y\succ _{1}w\sim _{1}z$
and $z\succ _{2}y\succ _{1}w\sim _{1}x:$ it can be quite easily shown that $%
\succcurlyeq _{1}$and $\succcurlyeq _{2}$ are unimodal -though of course not
strictly unimodal- \textit{and} violate the `Sequential Inclusion' property.
Furthermore, it can be quite easily checked that the foregoing preference
profile also fails to satisfy Indirect Sequential Inclusion (more details on
all of the above are available from the authors upon request).}

Finally, it should also be noticed that some of the results of the present
paper - notably, Lemma 1 - can be easily reproduced in a more general
setting e.g. in any median algebra (see Isbell (1980), Bandelt and Hedl\'{\i}%
kov\'{a} (1983)). It remains to be seen which of the other results, if any,
can also be lifted to the latter environment. This is however best left as a
possible topic for future research.

\section{Appendix 1: Proofs}

\begin{proof}[Proof of Claim 1]
(i) If $B_{\mathcal{X}}(x,z,y)$ then $x\wedge y\leqslant z\leqslant x\vee y$
. Since by definition $x\wedge y=y\wedge x$ and $x\vee y=y\vee x$ it
obviously follows $y\wedge x\leqslant z\leqslant y\vee x$ hence $B_{\mathcal{%
X}}(y,z,x)$ also holds.

(ii) Since by definition $x\wedge y\leqslant x\leqslant x\vee y$ and $%
x\wedge y\leqslant y\leqslant x\vee y$ hold for any $x,y\in X$, both $B_{%
\mathcal{X}}(x,x,y)$ and $B_{\mathcal{X}}(x,y,y)$ hold.

(iii) If $B_{\mathcal{X}}(x,y,x)$ then $x=x\wedge x\leqslant y\leqslant
x\vee x=x$ hence $y=x$.

(iv) If $B_{\mathcal{X}}(x,u,y),$ $B_{\mathcal{X}}(x,v,y)$ and $B_{\mathcal{X%
}}(u,z,v)$ then $x\wedge y\leqslant u\leqslant x\vee y$ , $x\wedge
y\leqslant v\leqslant x\vee y$ and $u\wedge v\leqslant z\leqslant u\vee v$.

Thus, by definition of $\wedge $ and $\vee $, $x\wedge y\leqslant u\wedge
v\leqslant x\vee y$ (that implies $x\wedge y\leqslant z$) and $x\wedge
y\leqslant u\vee v\leqslant x\vee y$ (that implies $z\leqslant x\vee y$). It
follows that $B_{\mathcal{X}}(x,z,y)$ as required;

(v) If $B_{\mathcal{X}}(x,y,z)$ and $B_{\mathcal{X}}(y,x,z)$ then $x\wedge
z\leqslant y\leqslant x\vee z$ \ and $y\wedge z\leqslant x\leqslant y\vee z$
hence

$x=x\vee (y\wedge z)=(x\wedge (y\vee z))\vee (y\wedge z)=(x\wedge y)\vee
(x\wedge z)\vee (y\wedge z)=$

$(x\wedge y)\vee (y\wedge z)\vee (x\wedge z)=(y\vee (x\wedge z))\vee
(x\wedge z)=y\vee (x\wedge z)=y$.
\end{proof}

\bigskip

\begin{proof}[Proof of Claim 2]
\textbf{\ }First, notice that $\succ _{x}$is, as suggested by notation,
asymmetric: indeed, suppose not i.e. $y\succ _{x}z$ and $z\succ _{x}y$ for
some $y,z\in X$. Then, in particular $y\in \lbrack x,z]$ and $z\in \lbrack
x,y]$ hence $y=z$ by antisymmetry of $B_{\mathcal{X}}$ as established above,
a contradiction. Next, observe that $\succ _{x}$is S-consistent: to check
this, assume that on the contrary there exist $y,z,z_{1},...,z_{k}$ such
that $y\in \lbrack x,z_{1}]\smallsetminus \left\{ z_{1}\right\} ,z_{1}\in
\lbrack x,z_{2}]\smallsetminus \left\{ z_{2}\right\} ,...,z_{k-1}\in \lbrack
x,z_{k}]\smallsetminus \left\{ z_{k}\right\} ,z_{k}\in \lbrack
x,z]\smallsetminus \left\{ z\right\} $ and $z\in \lbrack x,y]\smallsetminus
\left\{ y\right\} .$ Then, in particular,

$B_{\mathcal{X}}(x,y,z_{1})$, $B_{\mathcal{X}}(x,z_{1},z_{2})$, $B_{\mathcal{%
X}}(x,z_{2},z_{3})$, ....., $B_{\mathcal{X}}(x,z_{k-1},z_{k})$, $B_{\mathcal{%
X}}(x,z_{k},z)$, $B_{\mathcal{X}}(x,z,y).$

It follows, by closure and convexity of $B_{\mathcal{X}}$,

$[x,y]\subseteq \lbrack x,z_{1}]\subseteq \lbrack x,z_{2}]\subseteq
...\subseteq \lbrack x,z_{k}]\subseteq \lbrack x,z]\subseteq \lbrack x,y]$

hence $[x,y]=[x,z]$.

Therefore,

$B_{\mathcal{X}}(x,z,y)$ and $B_{\mathcal{X}}(x,y,z)$, hence by symmetry $B_{%
\mathcal{X}}(y,z,x)$ and $B_{\mathcal{X}}(z,y,x)$ and by antisymmetry $y=z$,
a contradiction since $z\in \lbrack x,y]\smallsetminus \left \{ y\right \} $%
: hence, S-consistency of $\succ _{x}$ is established.

It follows that Suzumura's Theorem applies and $\succ _{x}$admits a
non-trivial extension $\succcurlyeq _{x}^{\ast }$that is a total preorder.
Moreover, since $\succ _{x}$is asymmetric, $\succ _{x}\subseteq \succ
_{x}^{\ast }$hence by construction $\succcurlyeq _{x}^{\ast }$is locally
strictly unimodal as required.
\end{proof}

\textit{\ }

\begin{proof}[Proof of Lemma 1]
$(i)\Rightarrow (ii)$ Let $f$ be $B_{\mathcal{X}}$-monotonic with respect to 
$\mathcal{X}$. Now, consider any $\mathbf{\succcurlyeq }=(\succcurlyeq
_{j})_{j\in N}\in U_{\mathcal{X}}^{N}$ and any $i\in N$. By definition of $%
B_{\mathcal{X}}$-monotonicity $f(top(\succcurlyeq _{i}),x_{N\smallsetminus
\left \{ i\right \} })\in \lbrack top(\succcurlyeq
_{i}),f(x_{i},x_{N\smallsetminus \left \{ i\right \} })]$ for all $%
x_{N\smallsetminus \left \{ i\right \} }\in X^{N\smallsetminus \left \{
i\right \} }$ and $x_{i}\in X$. But then, since clearly

$top(\succcurlyeq _{i})\succcurlyeq _{i}f(top(\succcurlyeq
_{i}),x_{N\smallsetminus \left \{ i\right \} })$, either $f(top(\succcurlyeq
_{i}),x_{N\smallsetminus \left \{ i\right \} })=top(\succcurlyeq _{i})$ or $%
f(top(\succcurlyeq _{i}),x_{N\smallsetminus \left \{ i\right \}
})\succcurlyeq _{i}f(x_{i},x_{N\smallsetminus \left \{ i\right \} })$ by
unimodality of $\succcurlyeq _{i}$. Hence, $f(top(\succcurlyeq
_{i}),x_{N\smallsetminus \left \{ i\right \} })\succcurlyeq
_{i}f(x_{i},x_{N\smallsetminus \left \{ i\right \} })$ in any case. It
follows that $f$ is indeed strategy-proof on $U_{\mathcal{X}}^{N}$.

$(ii)\Rightarrow (i)$ Let us assume that $f:X^{N}\rightarrow X$ is \textit{%
not }$B_{\mathcal{X}}$-monotonic: thus, there exist $i\in N$, $x_{i}^{\prime
}\in X$ and $x_{N}=(x_{i})_{i\in N}\in X^{N}$ such that $f(x_{N})\notin
\lbrack x_{i},f(x_{i}^{\prime },x_{N\smallsetminus \left \{ i\right \} })]$.
Then, consider the total preorder $\succcurlyeq ^{\ast }$ on $X$ defined as
follows: $x_{i}=top(\succcurlyeq ^{\ast })$ and for all $y,z\in
X\smallsetminus \left \{ x_{i}\right \} $, $y\succcurlyeq ^{\ast }z$ if and
only if $(i)$ $\left \{ y,z\right \} \subseteq \lbrack x_{i},f(x_{i}^{\prime
},x_{N\smallsetminus \left \{ i\right \} })]\smallsetminus \left \{
x_{i}\right \} $ or $(ii)$ $y\in \lbrack x_{i},f(x_{i}^{\prime
},x_{N\smallsetminus \left \{ i\right \} })]\smallsetminus \left \{
x_{i}\right \} $ and $z\notin \lbrack x_{i},f(x_{i}^{\prime
},x_{N\smallsetminus \left \{ i\right \} })]$ or $(iii)$ $y\notin \lbrack
x_{i},f(x_{i}^{\prime },x_{N\smallsetminus \left \{ i\right \} })]$ and $%
z\notin \lbrack x_{i},f(x_{i}^{\prime },x_{N\smallsetminus \left \{
i\right
\} })]$. Clearly, by construction $\succcurlyeq ^{\ast }$consists
of three indifference classes with $\left \{ x_{i}\right \} $, $%
[x_{i},f(x_{i}^{\prime },x_{N\smallsetminus \left \{ i\right \}
})]\smallsetminus \left \{ x_{i}\right \} $ and $X\smallsetminus \lbrack
x_{i},f(x_{i}^{\prime },x_{N\smallsetminus \left \{ i\right \} })]$ as top,
medium and bottom indifference classes, respectively. Now, observe that $%
\succcurlyeq ^{\ast }\in $ $U_{\mathcal{X}}$. To check this statement, take
any $y,z,v\in X$ such that $y\neq z$ and $v\in \lbrack y,z]$ (if $y=z$ then $%
v=y=z$ and there is in fact nothing to prove). If $\left \{ y,z\right \}
\subseteq \lbrack x_{i},f(x_{i}^{\prime },x_{N\smallsetminus \left \{
i\right \} })]$ then $B_{\mathcal{X}}(x_{i},v,f(x_{i}^{\prime
},x_{N\smallsetminus \left \{ i\right
\} })$ by convexity of $B_{\mathcal{X}%
}$, i.e. $v\in \lbrack x_{i},f(x_{i}^{\prime },x_{N\smallsetminus \left \{
i\right \} })]$. $v$ by construction $x_{i}\wedge f(x_{i}^{\prime
},x_{N\smallsetminus \left \{ i\right \} })\leqslant y\wedge z\leqslant
v\leqslant y\vee z\leqslant x_{i}\vee f(x_{i}^{\prime },x_{N\smallsetminus
\left \{ i\right \} })$, i.e. $v\in \lbrack x_{i},f(x_{i}^{\prime
},x_{N\smallsetminus \left \{ i\right \} })] $. Assume without loss of
generality that $y\neq x_{i}$: it follows that $v\succcurlyeq ^{\ast }y$ by
definition of $\succcurlyeq ^{\ast }$. If on the contrary $\left \{
y,z\right \} \cap (X\smallsetminus \lbrack x_{i},f(x_{i}^{\prime
},x_{N\smallsetminus \left \{ i\right \} })])\neq \varnothing $ then clearly
by definition of $\succcurlyeq ^{\ast }$there exists $w\in \left \{
y,z\right \} $ such that $v\succcurlyeq ^{\ast }w$. Thus, $\succcurlyeq
^{\ast }\in $ $U_{\mathcal{X}}$ as claimed. Also, by assumption $f(x_{N})\in
X\smallsetminus \lbrack x_{i},f(x_{i}^{\prime },x_{N\smallsetminus \left \{
i\right \} })]$ whence by construction $f(x_{i}^{\prime },x_{N\smallsetminus
\left \{ i\right \} })\succ ^{\ast }f(x_{N})$. But then, $f$ is \textit{not }%
strategy-proof on $U_{\mathcal{X}}^{N}$ .

$(i)\Rightarrow (iii)$ Again, let $f$ be $B_{\mathcal{X}}$-monotonic with
respect to $\mathcal{X}$. Now, consider any $\mathbf{\succcurlyeq }%
=(\succcurlyeq _{j})_{j\in N}\in S_{\mathcal{X}}^{N}$ and any $i\in N$. By
definition of $B_{\mathcal{X}}$-monotonicity $f(top(\succcurlyeq
_{i}),x_{N\smallsetminus \left \{ i\right \} })\in \lbrack top(\succcurlyeq
_{i}),f(x_{i},x_{N\smallsetminus \left \{ i\right \} })]$ for all $%
x_{N\smallsetminus \left \{ i\right \} }\in X^{N\smallsetminus \left \{
i\right \} }$ and $x_{i}\in X$. But then, either $f(top(\succcurlyeq
_{i}),x_{N\smallsetminus \left \{ i\right \} })=f(x_{i},x_{N\smallsetminus
\left \{ i\right \} })$ or $f(top(\succcurlyeq _{i}),x_{N\smallsetminus
\left \{ i\right \} })\succ _{i}f(x_{i},x_{N\smallsetminus \left \{
i\right
\} }) $ by locally strict unimodality of $\succcurlyeq _{i}$.
Hence, $f(top(\succcurlyeq _{i}),x_{N\smallsetminus \left \{ i\right \}
})\succcurlyeq _{i}f(x_{i},x_{N\smallsetminus \left \{ i\right \} })$ in any
case. It follows that $f$ is indeed strategy-proof on $S_{\mathcal{X}}^{N}$.

$(iii)\Rightarrow (i)$ Let us assume that $f:X^{N}\rightarrow X$ is \textit{%
not }$B_{\mathcal{X}}$-monotonic: thus, there exist $i\in N$, $x_{i}^{\prime
}\in X$ and $x_{N}=(x_{i})_{i\in N}\in X^{N}$ such that $f(x_{N})\notin
\lbrack x_{i},f(x_{i}^{\prime },x_{N\smallsetminus \left \{ i\right \} })]$.
Then, consider a binary relation

$\succ ^{\prime }$on $X$ defined by the following clauses $(\alpha )$ $%
x_{i}=top(\succ ^{\prime })$ i.e. $x_{i}\succ ^{\prime }y$ for all $y\in
X\smallsetminus \left \{ x_{i}\right \} $; $(\beta )$ if $\left \{
y,z\right
\} \subseteq \lbrack x_{i},f(x_{i}^{\prime },x_{N\smallsetminus
\left \{ i\right \} })]\smallsetminus \left \{ x_{i}\right \} $ then $y\succ
^{\prime }z$ if and only if $z\in \lbrack x_{i},y]$ $\smallsetminus \left \{
y\right
\} $; $(\gamma )$ if $y\in \lbrack x_{i},f(x_{i}^{\prime
},x_{N\smallsetminus \left \{ i\right \} })]\smallsetminus \left \{
x_{i}\right \} $ and $z\notin \lbrack x_{i},f(x_{i}^{\prime
},x_{N\smallsetminus \left \{ i\right \} })]$ then $y\succ ^{\prime }z$ ; $%
(\delta )$ if $\left \{ y,z\right \} \subseteq X\smallsetminus \lbrack
x_{i},f(x_{i}^{\prime },x_{N\smallsetminus \left \{ i\right \} })]$ then $%
y\succ ^{\prime }z$ if and only if $z\in \lbrack x_{i},y]\smallsetminus
\left \{ y\right \} $.

Then, observe that $\succ ^{\prime }\cap \left\{ \lbrack
x_{i},f(x_{i}^{\prime },x_{N\smallsetminus \left\{ i\right\} })]\right\}
^{2} $ and $\succ ^{\prime }\cap \left\{ X\smallsetminus \lbrack
x_{i},f(x_{i}^{\prime },x_{N\smallsetminus \left\{ i\right\} })]\right\}
^{2} $ amount to the restrictions of $\succ _{x_{i}}$as defined above (see
Claim 2) to $[x_{i},f(x_{i}^{\prime },x_{N\smallsetminus \left\{ i\right\}
})]$ and $X\smallsetminus \lbrack x_{i},f(x_{i}^{\prime },x_{N\smallsetminus
\left\{ i\right\} })]$, respectively, hence by Claim 2 are S-consistent and
therefore admits non-trivial extensions to total preorders $\succcurlyeq
_{1}^{\prime }$and $\succcurlyeq _{2}^{\prime }$on their respective \textit{%
disjoint} domains. It follows that $\succcurlyeq ^{\prime }=\succcurlyeq
_{1}^{\prime }\oplus \succcurlyeq _{2}^{\prime }$ as defined by the rule $%
x\succcurlyeq ^{\prime }y$ if and only if $(x\succcurlyeq _{1}^{\prime }y$, $%
x\succcurlyeq _{2}^{\prime }y$ or $x\in \lbrack x_{i},f(x_{i}^{\prime
},x_{N\smallsetminus \left\{ i\right\} })]$ and $y\in X\smallsetminus
\lbrack x_{i},f(x_{i}^{\prime },x_{N\smallsetminus \left\{ i\right\} })]$)
is a total preorder on $X$. Moreover, $\succcurlyeq ^{\prime }$is locally
strictly unimodal by construction i.e. $\succcurlyeq ^{\prime }\in S_{%
\mathcal{X}}^{N}$. Also, by assumption $f(x_{N})\in X\smallsetminus \lbrack
x_{i},f(x_{i}^{\prime },x_{N\smallsetminus \left\{ i\right\} })]$ whence by
construction $f(x_{i}^{\prime },x_{N\smallsetminus \left\{ i\right\} })\succ
^{\prime }f(x_{N})$. But then, $f$ is \textit{not }strategy-proof on $S_{%
\mathcal{X}}^{N}$ .
\end{proof}

\begin{proof}[Proof of Lemma 2]
Take any $x_{N}\in X^{N}$. By definition of $B_{\mathcal{X}}$-monotonicity,
it suffices to show that for any $i\in N$ and $x_{i}^{\prime }\in X$, $\mu
(f,g,h)(x_{N})\in \lbrack x_{i},\mu (f,g,h)(x_{i}^{\prime
},x_{N\smallsetminus \left\{ i\right\} })]$. Indeed, by monotonicity of $%
f,g,h$ with respect to $\mathcal{X}$, $f(x_{N})\in \lbrack
x_{i},f(x_{i}^{\prime },x_{N\smallsetminus \left\{ i\right\} })]$, $%
g(x_{N})\in \lbrack x_{i},g(x_{i}^{\prime },x_{N\smallsetminus \left\{
i\right\} })]$, and $h(x_{N})\in \lbrack x_{i},h(x_{i}^{\prime
},x_{N\smallsetminus \left\{ i\right\} })]$.

A change of variables is in order here for the sake of convenience, namely $%
x_{f}=f(x_{N})$, $x_{f}^{\prime }=f(x_{i}^{\prime },x_{N\smallsetminus
\left\{ i\right\} })$, $x_{g}=g(x_{N})$, $x_{g}^{\prime }=g(x_{i}^{\prime
},x_{N\smallsetminus \left\{ i\right\} })$, $x_{h}=$ $h(x_{N})$, $%
x_{h}^{\prime }=h(x_{i}^{\prime },x_{N\smallsetminus \left\{ i\right\} })$,
whence $\mu (f,g,h)(x_{N})=\mu (x_{f},x_{g},x_{h})$, and $\mu
(f,g,h)(x_{i}^{\prime },x_{N\smallsetminus \left\{ i\right\} })=\mu
(x_{f}^{\prime },x_{g}^{\prime },x_{h}^{\prime })$. Thus, $x_{i}\wedge
x_{l}^{\prime }\leqslant x_{l}\leqslant x_{i}\vee x_{l}^{\prime }$, $l=f,g,h$%
, by hypothesis, while the thesis amounts to $x_{i}\wedge \mu (x_{f}^{\prime
},x_{g}^{\prime },x_{h}^{\prime })\leqslant \mu (x_{f},x_{g},x_{h})\leqslant
x_{i}\vee \mu (x_{f}^{\prime },x_{g}^{\prime },x_{h}^{\prime })$. Now, $\mu
(x_{f}^{\prime },x_{g}^{\prime },x_{h}^{\prime })=(x_{f}^{\prime }\wedge
x_{g}^{\prime })\vee (x_{g}^{\prime }\wedge x_{h}^{\prime })\vee
(x_{f}^{\prime }\wedge x_{h}^{\prime })$ hence by distributivity and the
basic latticial identities we get:%
\begin{eqnarray*}
&&x_{i}\wedge ((x_{f}^{\prime }\wedge x_{g}^{\prime })\vee (x_{g}^{\prime
}\wedge x_{h}^{\prime })\vee (x_{f}^{\prime }\wedge x_{h}^{\prime })) \\
&=&(x_{i}\wedge (x_{f}^{\prime }\wedge x_{g}^{\prime }))\vee (x_{i}\wedge
(x_{g}^{\prime }\wedge x_{h}^{\prime }))\vee (x_{i}\wedge (x_{f}^{\prime
}\wedge x_{h}^{\prime })) \\
&=&((x_{i}\wedge x_{f}^{\prime })\wedge (x_{i}\wedge x_{g}^{\prime }))\vee
((x_{i}\wedge x_{g}^{\prime })\wedge (x_{i}\wedge x_{h}^{\prime }))\vee
((x_{i}\wedge x_{f}^{\prime })\wedge (x_{i}\wedge x_{h}^{\prime }))\text{.}
\end{eqnarray*}

However, by hypothesis, distributivity and the basic latticial identities
again:%
\begin{eqnarray*}
&&((x_{i}\wedge x_{f}^{\prime })\wedge (x_{i}\wedge x_{g}^{\prime }))\vee
((x_{i}\wedge x_{g}^{\prime })\wedge (x_{i}\wedge x_{h}^{\prime }))\vee
((x_{i}\wedge x_{f}^{\prime })\wedge (x_{i}\wedge x_{h}^{\prime })) \\
&\leqslant &(x_{f}\wedge x_{g})\vee (x_{g}\wedge x_{h})\vee (x_{f}\wedge
x_{h})=\mu (x_{f},x_{g},x_{h}) \\
&\leqslant &((x_{i}\vee x_{f}^{\prime })\wedge (x_{i}\vee x_{g}^{\prime
}))\vee ((x_{i}\vee x_{g}^{\prime })\wedge (x_{i}\vee x_{h}^{\prime }))\vee
((x_{i}\vee x_{f}^{\prime })\wedge (x_{i}\vee x_{h}^{\prime })) \\
&=&(x_{i}\vee (x_{f}^{\prime }\wedge x_{g}^{\prime }))\vee (x_{i}\vee
(x_{g}^{\prime }\wedge x_{h}^{\prime }))\vee (x_{i}\vee (x_{f}^{\prime
}\wedge x_{h}^{\prime })) \\
&=&x_{i}\vee ((x_{f}^{\prime }\wedge x_{g}^{\prime })\vee (x_{g}^{\prime
}\wedge x_{h}^{\prime })\vee (x_{f}^{\prime }\wedge x_{h}^{\prime
}))=x_{i}\vee \mu (x_{f}^{\prime },x_{g}^{\prime },x_{h}^{\prime })
\end{eqnarray*}%
as required.
\end{proof}

\begin{proof}[Proof of Lemma 3]
Take any $x_{N\smallsetminus \left\{ 1\right\} }\in X^{N\smallsetminus
\left\{ 1\right\} }$ and consider $f_{x_{N\smallsetminus \left\{ 1\right\}
}}:X\rightarrow X$ as defined by the rule $f_{x_{N\smallsetminus \left\{
1\right\} }}(x_{1})=f(x_{1},x_{N\smallsetminus \left\{ 1\right\} })$ for all 
$x_{1}\in X$. \ Thus, by definition $f_{x_{N\smallsetminus \left\{ 1\right\}
}}$ is $B_{\mathcal{X}}$-monotonic with respect to $(X,\leqslant )$, i.e. $%
f_{x_{N\smallsetminus \left\{ 1\right\} }}(x)\in \lbrack x,$ $%
f_{x_{N\smallsetminus \left\{ 1\right\} }}(y)]$, namely 
\begin{equation*}
x\wedge f_{x_{N\smallsetminus \left\{ 1\right\} }}(y)\leqslant
f_{x_{N\smallsetminus \left\{ 1\right\} }}(x)\leqslant x\vee
f_{x_{N\smallsetminus \left\{ 1\right\} }}(y)\text{ for any }x,y\in X.
\end{equation*}%
In particular,%
\begin{eqnarray*}
\bot &=&\bot \wedge f_{x_{N\smallsetminus \left\{ 1\right\}
}}(x_{1})\leqslant f_{x_{N\smallsetminus \left\{ 1\right\} }}(\bot
)\leqslant \bot \vee f_{x_{N\smallsetminus \left\{ 1\right\}
}}(x_{1})=f_{x_{N\smallsetminus \left\{ 1\right\} }}(x_{1}), \\
f_{x_{N\smallsetminus \left\{ 1\right\} }}(x_{1}) &=&\top \wedge
f_{x_{N\smallsetminus \left\{ 1\right\} }}(x_{1})\leqslant
f_{x_{N\smallsetminus \left\{ 1\right\} }}(\top )\leqslant \top \vee
f_{x_{N\smallsetminus \left\{ 1\right\} }}(x_{1})=\top , \\
x_{1}\wedge f_{x_{N\smallsetminus \left\{ 1\right\} }}(\bot ) &\leqslant
&f_{x_{N\smallsetminus \left\{ 1\right\} }}(x_{1})\leqslant x_{1}\vee
f_{x_{N\smallsetminus \left\{ 1\right\} }}(\bot )\text{, and} \\
x_{1}\wedge f_{x_{N\smallsetminus \left\{ 1\right\} }}(\top ) &\leqslant
&f_{x_{N\smallsetminus \left\{ 1\right\} }}(x_{1})\leqslant x_{1}\vee
f_{x_{N\smallsetminus \left\{ 1\right\} }}(\top )\text{, for all }x_{1}\in X%
\text{.}
\end{eqnarray*}%
Now, take any $x_{1}\in X$ and consider $\mu (f_{x_{N\smallsetminus \left\{
1\right\} }}(\bot ),x_{1},f_{x_{N\smallsetminus \left\{ 1\right\} }}(\top ))$%
. By definition, and distributivity of $(X,\leqslant )$,%
\begin{eqnarray*}
&&\mu (f_{x_{N\smallsetminus \left\{ 1\right\} }}(\bot
),x_{1},f_{x_{N\smallsetminus \left\{ 1\right\} }}(\top )) \\
&=&(x_{1}\wedge f_{x_{N\smallsetminus \left\{ 1\right\} }}(\bot ))\vee
(x_{1}\wedge f_{x_{N\smallsetminus \left\{ 1\right\} }}(\top ))\vee
(f_{x_{N\smallsetminus \left\{ 1\right\} }}(\bot )\wedge
f_{x_{N\smallsetminus \left\{ 1\right\} }}(\top )) \\
&=&(x_{1}\vee f_{x_{N\smallsetminus \left\{ 1\right\} }}(\bot ))\wedge
(x_{1}\vee f_{x_{N\smallsetminus \left\{ 1\right\} }}(\top )\wedge
(f_{x_{N\smallsetminus \left\{ 1\right\} }}(\bot )\vee f_{x_{N\smallsetminus
\left\{ 1\right\} }}(\top ))\text{.}
\end{eqnarray*}

Since by $B_{\mathcal{X}}$-monotonicity, as observed above,%
\begin{eqnarray*}
x_{1}\wedge f_{x_{N\smallsetminus \left \{ 1\right \} }}(\bot ) &\leqslant
&f_{x_{N\smallsetminus \left \{ 1\right \} }}(x_{1})\text{, }x_{1}\wedge
f_{x_{N\smallsetminus \left \{ 1\right \} }}(\top )\leqslant
f_{x_{N\smallsetminus \left \{ 1\right \} }}(x_{1})\text{, and} \\
f_{x_{N\smallsetminus \left \{ 1\right \} }}(\bot )\wedge
f_{x_{N\smallsetminus \left \{ 1\right \} }}(\top ) &=&f_{x_{N\smallsetminus
\left \{ 1\right \} }}(\bot )\leqslant f_{x_{N\smallsetminus \left \{
1\right \} }}(x_{1})\text{,}
\end{eqnarray*}%
it follows that $\mu (f_{x_{N\smallsetminus \left \{ 1\right \} }}(\bot
),x_{1},f_{x_{N\smallsetminus \left \{ 1\right \} }}(\top ))\leqslant
f_{x_{N\smallsetminus \left \{ 1\right \} }}(x_{1})$. Similarly,%
\begin{eqnarray*}
f_{x_{N\smallsetminus \left \{ 1\right \} }}(x_{1}) &\leqslant &x_{1}\vee
f_{x_{N\smallsetminus \left \{ 1\right \} }}(\bot ),f_{x_{N\smallsetminus
\left \{ 1\right \} }}(x_{1})\leqslant x_{1}\vee f_{x_{N\smallsetminus \left
\{ 1\right \} }}(\top )\text{, and} \\
f_{x_{N\smallsetminus \left \{ 1\right \} }}(x_{1}) &\leqslant
&f_{x_{N\smallsetminus \left \{ 1\right \} }}(\top )=f_{x_{N\smallsetminus
\left \{ 1\right \} }}(\bot )\vee f_{x_{N\smallsetminus \left \{ i\right \}
}}(\top )\text{.}
\end{eqnarray*}%
It follows that $f_{x_{N\smallsetminus \left \{ 1\right \}
}}(x_{1})\leqslant \mu (f_{x_{N\smallsetminus \left \{ 1\right \} }}(\bot
),x_{1},f_{x_{N\smallsetminus \left \{ 1\right \} }}(\top ))$ as well, whence%
\begin{eqnarray*}
f_{x_{N\smallsetminus \left \{ 1\right \} }}(x_{1}) &=&\mu
(f_{x_{N\smallsetminus \left \{ 1\right \} }}(\bot
),x_{1},f_{x_{N\smallsetminus \left \{ 1\right \} }}(\top ))=\mu
(f_{x_{N\smallsetminus \left \{ 1\right \} }}(\bot ),\pi
_{1}(x_{1}),f_{x_{N\smallsetminus \left \{ 1\right \} }}(\top )) \\
&=&\mu (f(\bot ,x_{N\smallsetminus \left \{ 1\right \} }),\pi
_{1}(x_{1}),f(\top ,x_{N\smallsetminus \left \{ 1\right \} }))\text{, i.e. }
\\
f_{x_{N\smallsetminus \left \{ 1\right \} }} &=&\mu (f(\bot
,x_{N\smallsetminus \left \{ 1\right \} }),\pi _{1},f(\top
,x_{N\smallsetminus \left \{ 1\right \} })).
\end{eqnarray*}%
Thus, for all $x_{1}\in X$, $f_{x_{N\smallsetminus \left \{ 1\right \}
}}(x_{1})$ is the \textit{first} term of the nested sequence of medians that
provides the run of the median tree-automaton $\mathcal{A}_{\mu }^{I,\lambda
}$ as initialized with ballot profile $x_{N}$ and applied to the finite $%
(\Sigma ^{\mu },I)$-tree $T=T(x_{N}$,$\left \{ f(x^{\ast })\right \}
_{x^{\ast }\in \left \{ \bot ,\top \right \} ^{N}})$ with terminal nodes
suitably labelled by projections of $x_{N}$ and elements of $\left \{
f(x^{\ast })\right \} _{x^{\ast }\in \left \{ \bot ,\top \right \} ^{N}}$.

Next, consider $f_{x_{N\smallsetminus \left\{ 1,2\right\}
}}:X^{2}\rightarrow X$ as defined by the following rule: for all $%
x_{1},x_{2}\in X$,%
\begin{eqnarray*}
f_{x_{N\smallsetminus \left\{ 1,2\right\} }}(x_{1},x_{2})
&=&f(x_{1},x_{2},x_{N\smallsetminus \left\{ 1,2\right\} })=\mu (f(\bot
,x_{N\smallsetminus \left\{ 1\right\} }),\pi _{1}(x_{1}),f(\top
,x_{N\smallsetminus \left\{ 1\right\} })) \\
&=&\mu (f(\bot ,x_{2},x_{N\smallsetminus \left\{ 1,2\right\} }),\pi
_{1}(x_{1}),f(\top ,x_{2},x_{N\smallsetminus \left\{ 1,2\right\} }))\text{.}
\end{eqnarray*}%
By repeating the previous argument of this proof as applied to both $f(\bot
,x_{2},x_{N\smallsetminus \left\{ 1,2\right\} })$ and $f(\top
,x_{2},x_{N\smallsetminus \left\{ 1,2\right\} })$, it follows that:%
\begin{eqnarray*}
f_{x_{N\smallsetminus \left\{ 1,2\right\} }}(x_{1},x_{2}) &=&\mu (\mu
(f(\bot ,\bot ,x_{N\smallsetminus \left\{ 1,2\right\} }),\pi
_{2}(x_{2}),f(\bot ,\top ,x_{N\smallsetminus \left\{ 1,2\right\} })), \\
&&\pi _{1}(x_{1}),\mu (f(\top ,\bot ,x_{N\smallsetminus \left\{ 1,2\right\}
}),\pi _{2}(x_{2}),f(\top ,\top ,x_{N\smallsetminus \left\{ 1,2\right\} })))%
\text{,}
\end{eqnarray*}%
i.e. $f_{x_{N\smallsetminus \left\{ 1,2\right\} }}$ is the \textit{fourth }%
term of the nested sequence of medians that provides the run of the median
tree-automaton $\mathcal{A}_{\mu }^{I,\lambda }$ as initialized with ballot
profile $x_{N}$ and applied to the finite $(\Sigma ^{\mu },I)$-tree $%
T=T(x_{N}$,$\left\{ f(x^{\ast })\right\} _{x^{\ast }\in \left\{ \bot ,\top
\right\} ^{N}})$ \ as mentioned above. Repeated iteration of the very same
argument establishes that, for all $x_{N}\in X^{N}$, $f(x_{N})=\mathcal{A}%
_{\mu }^{I,\lambda }(T)$ i.e. $f(x_{N})$ is the value of the behaviour of $%
\mathcal{A}_{\mu }^{I,\lambda }$ at $T$.
\end{proof}

\begin{proof}[Proof of Theorem 1]
$(i)\Longleftrightarrow (ii)$ It follows from Lemma 1.

$(i)\Longleftrightarrow (iii)$ It also follows from Lemma 1.

$(i)\Longrightarrow (iv)$ Immediate from Lemma 3.

$(iv)\Longrightarrow (i)$ It follows immediately from the definition of
l-median TA-representation, from the observation that projections and
constants induce $B_{\mathcal{X}}$ -monotonic voting rules, and from Lemma 2.

$(iv)\Longrightarrow (v)$ Suppose $f$ \ has a $\Sigma ^{\mu }$-tree automata
representation. Then, by construction, there exists a set of at most $%
2^{|N|} $ distinct elements namely

$f[\left\{ \bot ,\top \right\} ^{N}]=\left\{ f(z_{N}):z_{N}\in \left\{ \bot
,\top \right\} ^{N}\right\} $ such that $f(x_{N})$ is the output of an
l-median tree automaton with input set $I(f(x_{N})):=f[\left\{ \bot ,\top
\right\} ^{N}]\cup \left\{ x_{i}:i\in N\right\} $. But then, by definition
of the output computation rule of an l-median tree automaton it is
immediately checked that $f(x_{N})$ is the join of a finite set of meets $%
\wedge _{j=1}^{k}z_{j}$, $k\leq 2^{|N|}$, $z_{i}\in I(f(x_{N}))$, $i=1,...,k$%
. Hence, by positing

$\ y_{S}[\wedge _{j=1}^{k}z_{j}]=\wedge \left( z:z=z_{j}\text{ for some }%
j=1,...,k\text{ and }z_{j}\neq x_{i}\text{ for all }i\in N\right) $

where $S[\wedge _{j=1}^{k}z_{j}]:=\left \{ i\in N:x_{i}=z_{h}\text{ for some 
}h=1,...,k\right \} $, and

$y_{S}^{\ast }=\vee (y_{S}[\wedge _{j=1}^{k}z_{j}]$ such that $S[\wedge
_{j=1}^{k}z_{j}]=S)$,

and noticing that $y_{S}$ thus defined does not depend on $x_{N}$,

it follows that there exists a nonempty $\mathcal{C\subseteq P}(N)$ such that

$f(x_{N})=\vee _{S\in \mathcal{C}}((\wedge _{i\in S}x_{i})\wedge y_{S}^{\ast
})$.

Moreover, by putting $y_{T}^{\ast }=\bot $ for any $T\in \mathcal{P(}%
N)\smallsetminus \mathcal{C}$, it also follows that

$f(x_{N})=\vee _{S\subseteq N}((\wedge _{i\in S}x_{i})\wedge y_{S}^{\ast })$

hence $f$ is indeed a generalized weak committee voting rule.

$(v)\Longrightarrow (i)$ Let $f:X^{N}\rightarrow X$ be a generalized weak
committee voting rule i.e. there exists an order filter $\mathcal{F}$ of $(%
\mathcal{P}(N),\subseteq )$ such that $f(x_{N})=\vee _{S\in \mathcal{F}%
}((\wedge _{i\in S}x_{i})\wedge y_{S}^{\ast })$ for all $x_{N}\in X^{N}$.
Then, observe that -for any $x,y\in X$- $x\wedge y=\mu (x,y,\bot )$ and $%
x\vee y=\mu (x,y,\top )$. Hence by repeated application of Lemma 2 it
follows that $f$ is $B_{\mathcal{X}}$ -monotonic.
\end{proof}

\begin{proof}[Proof of Theorem 2]
$(i)$ Take restricted voting rule $f^{\prime }$ as introduced in Remark 1
above, where it was also shown that $f^{\prime }$ is strategy-proof on $%
D^{2}\times U_{\mathcal{X}}^{N\smallsetminus \left \{ 1,2\right \} }$ and on 
$(D^{\prime })^{2}\times S_{\mathcal{X}}^{N\smallsetminus \left \{
1,2\right
\} }$.

Now, to address the strategy-proofness issue on $D^{2}\times U_{\mathcal{X}%
}^{N\smallsetminus \left \{ 1,2\right \} }$ consider any preference profile $%
(\succcurlyeq _{i})_{i\in N}$ such that $\succcurlyeq _{1}=\succcurlyeq
^{\prime }$ and $\succcurlyeq _{2}=\succcurlyeq $ hence $top(\succcurlyeq
_{1})=d$, $top(\succcurlyeq _{2})=a$. Then, for any $x_{N\smallsetminus
\left \{ 1,2\right \} }\in X^{N\smallsetminus \left \{ 1,2\right \} }$, both%
\begin{eqnarray*}
f^{\prime }(a,d,x_{N\smallsetminus \left \{ 1,2\right \} }) &\succ
&_{1}f^{\prime }(top(\succcurlyeq _{1}),top(\succcurlyeq
_{2}),x_{N\smallsetminus \left \{ 1,2\right \} })\text{ and} \\
f^{\prime }(a,d,x_{N\smallsetminus \left \{ 1,2\right \} }) &\succ
&_{2}f^{\prime }(top(\succcurlyeq _{1}),top(\succcurlyeq
_{2}),x_{N\smallsetminus \left \{ 1,2\right \} })\text{,}
\end{eqnarray*}%
it follows that coalition $\left \{ 1,2\right \} $ can manipulate the
outcome at $(\succcurlyeq _{i})_{i\in N}$ namely $f^{\prime }$ is \textit{%
not }coalitionally strategy-proof on $D^{2}\times U_{\mathcal{X}%
}^{N\smallsetminus \left \{ 1,2\right \} }$. Strategy-proofness (and failure
of coalitional strategy-proofness) of $f^{\prime }$ on $(D^{\prime
})^{2}\times S_{\mathcal{X}}^{N\smallsetminus \left \{ 1,2\right \} }$ is
proved in a similar way by replacing $\succcurlyeq $ and $\succcurlyeq
^{\prime }$ with $\succcurlyeq ^{\prime \prime }$ and $\ \succcurlyeq
^{\prime \prime \prime }$, respectively.

$(ii)$ Let us assume without loss of generality that $|X|=4$ and let $%
X=\left \{ a,b,c,d\right \} $ and $\Delta _{X}=\left \{ (x,x):x\in
X\right
\} $. Next, define $\leqslant ^{\ast \ast }=\left \{
(a,b),(a,c),(a,d),(b,d),(c,d)\right \} \cup \Delta _{X}$.

It is easily checked that $\mathcal{X}^{\ast \ast }=(X,\leqslant ^{\ast \ast
})$ is the Boolean lattice $2^{2}$ with $a=\top $, $d=\bot $.

Now, define the family $\left \{ f(x^{\ast })\right \} _{x^{\ast }\in
\left
\{ \bot ,\top \right \} ^{N}}$ as follows: for all $%
x_{N\smallsetminus \left
\{ 1,2\right \} }\in \left \{ \bot ,\top \right \}
^{N\smallsetminus \left
\{ 1,2\right \} }$%
\begin{equation*}
f(a,a,x_{N\smallsetminus \left \{ 1,2\right \} })=a,\text{ }%
f(d,d,x_{N\smallsetminus \left \{ 1,2\right \} })=d,\text{ }%
f(a,d,x_{N\smallsetminus \left \{ 1,2\right \} })=b,\text{ }%
f(d,a,x_{N\smallsetminus \left \{ 1,2\right \} })=c\text{.}
\end{equation*}

Then, consider the nested sequence of medians that provides the run of the
median tree-automaton $\mathcal{A}_{\mu }^{I,\lambda }$ as initialized with
ballot profile $x_{N}$ and applied to the finite $(\Sigma ^{\mu },I)$-tree $%
T=T(x_{N}$,$\left\{ f(x^{\ast })\right\} _{x^{\ast }\in \left\{ \bot ,\top
\right\} ^{N}})$ with terminal nodes suitably labelled by projections of $%
x_{N}$ and elements of $\left\{ f(x^{\ast })\right\} _{x^{\ast }\in \left\{
\bot ,\top \right\} ^{N}}$ as defined above (notice that $f$ is by
construction an extension of $f^{\prime }$ to $X^{N}$ as mentioned above
under part $(i)$ of the present proof). A few simple if tedious calculations
immediately establish that for all

$x_{N\smallsetminus \left \{ 1,2\right \} }\in X^{N\smallsetminus \left \{
1,2\right \} }$:%
\begin{eqnarray*}
f(a,c,x_{N\smallsetminus \left \{ 1,2\right \} })
&=&f(b,a,x_{N\smallsetminus \left \{ 1,2\right \}
})=f(b,c,x_{N\smallsetminus \left \{ 1,2\right \} })=a, \\
f(b,b,x_{N\smallsetminus \left \{ 1,2\right \} })
&=&f(a,b,x_{N\smallsetminus \left \{ 1,2\right \}
})=f(b,d,x_{N\smallsetminus \left \{ 1,2\right \} })=b, \\
f(c,c,x_{N\smallsetminus \left \{ 1,2\right \} })
&=&f(c,a,x_{N\smallsetminus \left \{ 1,2\right \}
})=f(d,c,x_{N\smallsetminus \left \{ 1,2\right \} })=c, \\
f(c,d,x_{N\smallsetminus \left \{ 1,2\right \} })
&=&f(d,c,x_{N\smallsetminus \left \{ 1,2\right \}
})=f(c,b,x_{N\smallsetminus \left \{ 1,2\right \} })=d.
\end{eqnarray*}%
By construction, and in view of Lemma 3 above, $f$ is $B_{\mathcal{X}^{\ast
\ast }}$- monotonic. Therefore, by Lemma 1, $f$ is also strategy-proof on $%
U_{\mathcal{X}^{\ast \ast }}^{N}$.

Now, take%
\begin{eqnarray*}
&\succcurlyeq &=\left \{ (a,b),(a,c),(a,d),(b,c),(b,d),(c,d),(d,c)\right \}
\cup \Delta _{X}, \\
&\succcurlyeq &^{^{\prime }}=\left \{
(d,b),(d,c),(d,a),(b,c),(b,a),(c,a),(a,c)\right \} \cup \Delta _{X},
\end{eqnarray*}

as defined in Remark 2 above.

\ First, observe that both $\succcurlyeq $ and $\succcurlyeq ^{\prime }$are
in $U_{\mathcal{X}^{\ast \ast }}^{N}$, i.e. are unimodal with respect to $%
\mathcal{X}^{\ast \ast }$: indeed, $top(\succcurlyeq )=a$, $top(\succcurlyeq
^{\prime })=d$ and it is immediately seen that%
\begin{eqnarray*}
B_{\mathcal{X}}(X, &\leqslant &^{\ast \ast })=\left \{ 
\begin{array}{c}
(a,b,d),(a,c,d),(b,a,c),(b,d,c),(d,b,a), \\ 
(d,c,a),(c,a,b),(c,d,b)%
\end{array}%
\right \} \cup \\
&&\cup \left \{ (x,y,z)\in X^{3}:x=y\text{ or }z=y\right \} \text{.}
\end{eqnarray*}%
But then, since $\left \{ (b,d),(c,d),(a,b),(d,c)\right \} \cup \Delta _{X}$%
\ is a subrelation of $\succcurlyeq $ and $\left \{
(b,a),(c,a),(a,c),(d,c)\right \} \cup \Delta _{X}$ is a subrelation of $%
\succcurlyeq ^{\prime }$, it follows that $\succcurlyeq $ and $\succcurlyeq
^{\prime }$are also unimodal with respect to $\mathcal{X}^{\ast \ast }$.
Now, take any preference profile $(\succcurlyeq _{i})_{i\in N}$ such that $%
\succcurlyeq _{1}=\succcurlyeq ^{\prime }$ and $\succcurlyeq
_{2}=\succcurlyeq $, hence $top(\succcurlyeq _{1})=d$, $top(\succcurlyeq
_{2})=a$. Then, for any $x_{N\smallsetminus \left \{ 1,2\right \} }\in
X^{N\smallsetminus \left \{ 1,2\right \} }$, both $f(a,d,x_{N\smallsetminus
\left \{ 1,2\right \} })\succ _{1}f(top(\succcurlyeq _{1}),top(\succcurlyeq
_{2}),x_{N\smallsetminus \left \{ 1,2\right \} })$ and $f(a,d,x_{N%
\smallsetminus \left \{ 1,2\right \} })\succ _{2}f(top(\succcurlyeq
_{1}),top(\succcurlyeq _{2}),x_{N\smallsetminus \left \{ 1,2\right \} })$:
it follows that, again, coalition $\left \{ 1,2\right \} $ can manipulate
the outcome at $(\succcurlyeq _{i})_{i\in N}$ namely $f$ is \textit{not }%
coalitionally strategy-proof.

Again, strategy-proofness and failure of coalitional strategy-proofness of $%
f $ on $S_{\mathcal{X}^{\ast \ast }}^{N}$ follows from the very same
argument, by positing $\succcurlyeq _{1}=\succcurlyeq ^{\prime \prime }$ and 
$\succcurlyeq _{2}=\succcurlyeq ^{\prime \prime \prime }$.
\end{proof}

\begin{proof}[Proof of Corollary 1]
$(i)\Longrightarrow (ii)$ It follows immediately from Theorem 2 (ii) above;

$(ii)\Longrightarrow (i)$ For the case concerning $U_{\mathcal{Y}}^{N}$, the
statement follows from a straightforward extension and adaptation of the
proof of Proposition 4 of Danilov (1994) concerning voting rules on unimodal
domains of \textit{linear orders }in \textit{undirected bounded trees }%
(details available from the authors upon request), and is indeed already
stated without explicit proof in Moulin (1980). As far as $S_{\mathcal{Y}%
}^{N}$ is concerned, the statement follows e.g. from Theorem 2 and
Proposition 3 of Barber\`{a}, Berga and Moreno (2010).
\end{proof}

\begin{proof}[Proof of Thorem 3]
Let us assume that on the contrary there exists a voting rule $%
f:X^{N}\rightarrow X$ which is anonymous, locally JI-neutral on $Y$, locally
sovereign on $Y$, and coalitionally strategy-proof on $U_{\mathcal{X}}^{N}$
(on $S_{\mathcal{X}}^{N}$ , respectively). By Theorem 1, it follows that
there exists an order filter $\mathcal{F}$ of $(\mathcal{P}(N),\subseteq )$
such that

$f(x_{N})=\vee _{S\in \mathcal{F}}((\wedge _{i\in S}x_{i})\wedge y_{S}^{\ast
})$ for all $x_{N}\in X^{N}$.

To begin with, observe that coalitional strategy-proofness and local
sovereignty on $Y$ jointly imply local idempotence on $Y$ (indeed, suppose
there exists $u\in Y$, $u\neq f(u^{N})$; of course, by local sovereignty
there exists $x_{N}\in X^{N}$ such that $f(x_{N})=u.$ But then $f$ is
coalitionally manipulable at any preference profile $(\succcurlyeq
_{i})_{i\in N}\in U_{\mathcal{X}}^{N}$ ($(\succcurlyeq _{i})_{i\in N}\in S_{%
\mathcal{X}}^{N}$ , respectively) such that $top(\succcurlyeq _{i})=u$ for
all $i\in N$, a contradiction).

Next, for any $u\in Y$ denote by $S_{u}$ the set of all minimal coalitions $%
T\in \mathcal{F}$ such that $u\leqslant f(u^{T},w^{N\smallsetminus T})$ for
all $w^{N\smallsetminus T}\in X^{N\smallsetminus T}$. By local idempotence
of $f$ on $Y$ , $S_{u}\neq \varnothing $. By anonymity of $f$, $%
|T|=|T^{\prime }|=n_{u}$ for all $T,T^{\prime }\in S_{u}$, and $y_{S}^{\ast
}=y_{S^{\prime }}^{\ast }=y_{s}^{\ast }$ for any $S,S^{\prime }\in \mathcal{F%
}$ such that $|S|=|S^{\prime }|=s$. Moreover, since by Theorem 1 coalitional
strategy-proofness entails in particular $B_{\mathcal{X}}$-monotonicity, it
also follows -by definition of $B_{\mathcal{X}}$-monotonicity- that for any $%
i\in N\smallsetminus T$

$u=u\wedge f(u^{T},w^{N\smallsetminus T})\leqslant f((u^{T\cup \left \{
i\right \} },w^{N\smallsetminus (T\cup \left \{ i\right \} )})\leqslant
u\vee f(u^{T},w^{N\smallsetminus T})$

whence, by repeated application of that argument

$u\leqslant f(u^{T^{\prime }},w^{N\smallsetminus T^{\prime }})$ for any $%
T^{\prime }\subseteq N$ such that $|T^{\prime }|\geq n_{u}$.

Also, by local JI-neutrality on $Y$ of $f$, $n_{x}=n_{z}=q$.

Four cases are to be distinguished according to the sign of $(q-n/2)$ and
the parity of $n$.

($\alpha $): Let us first suppose that $q\leq n/2$.

Then, in order to address the unimodal case consider the following triple of
preference relations:

$\succcurlyeq ^{\ast }:=[x\succ ^{\ast }0\succ ^{\ast }x\vee z\sim ^{\ast
}z\sim ^{\ast }w$ for all $w\in X\smallsetminus Y]$,

$\succcurlyeq ^{\ast \ast }:=[z\succ ^{\ast \ast }0\succ ^{\ast \ast }x\vee
z\sim ^{\ast \ast }x\sim ^{\ast \ast }w$ for all $w\in X\smallsetminus Y]$,

$\succcurlyeq ^{\ast \ast \ast }:=[0\succ ^{\ast \ast \ast }x\sim ^{\ast
\ast \ast }z\sim ^{\ast \ast \ast }x\vee z\sim ^{\ast \ast \ast }w$ for all $%
w\in X\smallsetminus Y]$.

Notice that by construction such preferences are unimodal with respect to $%
\mathcal{X}$, i.e. $\left \{ \succcurlyeq ^{\ast },\succcurlyeq ^{\ast \ast
},\succcurlyeq ^{\ast \ast \ast }\right \} \subseteq U_{\mathcal{X}}^{N}$.

Two subcases are distinguished according to the parity of $n$, namely

$(i)$ $n=2k+1$ for some positive integer $k$, and $\ (ii)\ n=2k$ \ for some
positive integer $k$.

If ($\alpha (i)$) obtains then take preference profile

$\mathbf{\succcurlyeq }_{[O]}=((\succcurlyeq _{i}^{\ast })_{i\in \left \{
1,...,k\right \} },(\succcurlyeq _{i}^{\ast \ast })_{i\in \left \{
k+1,...,2k\right \} },\succcurlyeq _{2k+1}^{\ast \ast \ast })$

and compute $f(y_{N})=\vee _{S\in \mathcal{F}}((\wedge _{i\in S}y_{i})\wedge
y_{s}^{\ast })$

where $y_{N}=top(\mathbf{\succcurlyeq }_{[O]})$ i.e. $y_{i}=x$ for all $i\in
\left \{ 1,...,k\right \} $, $y_{i}=z$ for all $i\in \left \{
k+1,...,2n\right \} $, and $y_{2k+1}=0$.

By construction, $f(y_{N})$ is the l.u.b. of a nonempty family $\mathcal{T}$
of terms belonging to some of the following jointly exhaustive, partially
overlapping classes:

$T_{1}=\left \{ \wedge _{j\in J}v_{j}:J\text{ \ is a finite set }J\text{ and
there exists }j\in J\text{ such that }v_{j}=0\right \} ,$

$T_{2}=\left \{ \wedge _{j\in J}v_{j}:J\text{ \ is\ a finite set }J\text{
and there exist }j,h\in J\text{ such that }v_{j}=x\text{ and }%
v_{h}=z\right
\} ,$

$T_{3}=\left \{ \wedge _{j\in J}v_{j}:\text{ }J\text{ is a finite set and
there exists }J^{\prime }\subseteq J\text{ such that }|J^{\prime }|\geq q%
\text{ and }v_{j}=x\text{ for all }j\in J^{\prime }\right \} ,$

$T_{4}=\left \{ \wedge _{j\in J}v_{j}:\text{ }J\text{ is a finite set and
there exists }J^{\prime }\subseteq J\text{ such that }|J^{\prime }|\geq q%
\text{ and }v_{j}=z\text{ for all }j\in J^{\prime }\right \} $.

Moreover, $t=\wedge _{j\in J}v_{j}=0$ for all $t\in T_{1}\cup T_{2}$ hence,
by construction, $T_{3}\cap \mathcal{T}\neq \varnothing \neq T_{4}\cap 
\mathcal{T}$. On the other hand, $t_{3}\geqslant x$ and $t_{4}\geqslant z$
for any $t_{3}\in T_{3}$ and $t_{4}\in T_{4}$.

It follows that \ $f(y_{N})\geqslant x\vee z$.

If ($\alpha (ii)$) obtains then take preference profile

$\mathbf{\succcurlyeq }_{[E]}=((\succcurlyeq _{i}^{\ast })_{i\in \left \{
1,...,k\right \} },(\succcurlyeq _{i}^{\ast \ast })_{i\in \left \{
k+1,...,2k\right \} }),$

and compute $f(y_{N}^{\prime })=\vee _{S\in \mathcal{F}}((\wedge _{i\in
S}y_{i}^{\prime })\wedge y_{s}^{\ast })$,

where $y_{N}^{\prime }=top(\mathbf{\succcurlyeq }_{[E]})$ i.e. $%
y_{i}^{\prime }=x$ for all $i\in \left \{ 1,...,k\right \} $, and $%
y_{i}^{\prime }=z$ for all $i\in \left \{ k+1,...,2n\right \} $,

Again, $f(y_{N}^{\prime })$ is the l.u.b. of a nonempty family $\mathcal{T}$
of terms belonging to some of the following jointly exhaustive, partially
overlapping classes:

$T_{1}^{\prime }=\left \{ \wedge _{j\in J}v_{j}:J\text{ \ is\ a finite set }J%
\text{ and there exist }j,h\in J\text{ such that }v_{j}=x\text{ and }%
v_{h}=z\right \} ,$

$T_{2}^{\prime }=\left \{ \wedge _{j\in J}v_{j}:\text{ }J\text{ is a finite
set and there exists }J^{\prime }\subseteq J\text{ such that }|J^{\prime
}|\geq q\text{ and }v_{j}=x\text{ for all }j\in J^{\prime }\right \} ,$

$T_{3}^{\prime }=\left \{ \wedge _{j\in J}v_{j}:\text{ }J\text{ is a finite
set and there exists }J^{\prime }\subseteq J\text{ such that }|J^{\prime
}|\geq q\text{ and }v_{j}=z\text{ for all }j\in J^{\prime }\right \} $.

Moreover, $t=\wedge _{j\in J}v_{j}=0$ for all $t\in T_{1}^{\prime }$ hence,
by construction, $T_{2}^{\prime }\cap \mathcal{T}\neq \varnothing \neq
T_{3}^{\prime }\cap \mathcal{T}$. On the other hand, $t_{2}\geqslant x$ and $%
t_{3}\geqslant z$ for any $t_{2}\in T_{2}^{\prime }$ and $t_{3}\in
T_{3}^{\prime }$.

It follows, again, that $\ f(y_{N}^{\prime })\geqslant x\vee z$.

Now, take $u_{N}\in X^{N}$ with $u_{i}=0$ for all $i\in N$: by local
idempotence, $f(u_{N})=0.$

Thus, if $n=2k+1$, $f((u_{i}=0)_{i\in N\smallsetminus \left \{ 2k+1\right \}
},y_{2k+1}=0)=f(u_{N})\succ _{i}f(y_{N})$ for all $i\in N\smallsetminus
\left \{ 2k+1\right \} $.

Similarly, if $n=2k,$ then $f(u_{N})$ $\succ _{i}f(y_{N})$ for all $i\in N$.
Hence, $f$ is coalitionally manipulable at unimodal preference profile $%
\mathbf{\succcurlyeq }^{[O]}$ (at unimodal preference profile $\mathbf{%
\succcurlyeq }^{[E]}$, respectively), a contradiction.

The locally strictly unimodal case can be addressed precisely by the same
argument, provided preference profile $(\succcurlyeq ^{\ast },\succcurlyeq
^{\ast \ast },\succcurlyeq ^{\ast \ast \ast })$ is replaced by any locally
strictly unimodal preference profile $(\succcurlyeq ^{\prime },\succcurlyeq
^{\prime \prime },\succcurlyeq ^{\prime \prime \prime })$ such that

$\succcurlyeq ^{\prime }:=[x\succ ^{\prime }0\succ ^{\prime }x\vee z\succ
^{\prime }z\succ w$ for all $w\in X\smallsetminus Y]$,

$\succcurlyeq ^{\prime \prime }:=[z\succ ^{\prime \prime }0\succ ^{\prime
\prime }x\vee z\succ ^{\prime \prime }x\succ ^{\prime \prime }w$ for all $%
w\in X\smallsetminus Y]$,

$\succcurlyeq ^{\prime \prime \prime }:=[0\succ ^{\prime \prime \prime
}x\succ ^{\prime \prime \prime }z\succ ^{\prime \prime \prime }x\vee z\succ
^{\prime \prime \prime }w$ for all $w\in X\smallsetminus Y]$.

($\beta $) Let us now assume that, on the contrary, $q>(n/2).$

Then, consider the following triple of preference relations:

$\succcurlyeq ^{\circ }:=[x\succ ^{\circ }x\vee z\succ ^{\circ }0\sim
^{\circ }z\sim ^{\circ }w$ for all $w\in X\smallsetminus Y]$,

$\succcurlyeq ^{\circ \circ }:=[z\succ ^{\circ \circ }x\vee z\succ ^{\circ
\circ }0\sim ^{\circ \circ }x\sim ^{\circ \circ }w$ for all $w\in
X\smallsetminus Y]$,

$\succcurlyeq ^{\circ \circ \circ }:=[0\succ ^{\circ \circ \circ }x\sim
^{\circ \circ \circ }z\sim ^{\circ \circ \circ }x\vee z\sim ^{\circ \circ
\circ }w$ for all $w\in X\smallsetminus Y]$.

Notice that by construction such preferences are unimodal with respect to $%
\mathcal{X}$, i.e. $\left \{ \succcurlyeq ^{\circ },\succcurlyeq ^{\circ
\circ },\succcurlyeq ^{\prime }\right \} \subseteq U_{\mathcal{X}}$.

Two subcases are distinguished again according to the parity of $n$, namely

$(i)$ $n=2k+1$ for some positive integer $k$, and $\ (ii)\ n=2k$ \ for some
positive integer $k$.

If ($\beta (i)$) obtains, then take preference profile

$\mathbf{\succcurlyeq }_{[O]}^{\circ }=((\succcurlyeq _{i}^{\circ })_{i\in
\left \{ 1,...,k\right \} },(\succcurlyeq _{i}^{\circ \circ })_{i\in
\left
\{ k+1,...,2k\right \} },\succcurlyeq _{n}^{\circ \circ \circ })$

and compute $f(w_{N})=\vee _{S\in \mathcal{F}}((\wedge _{i\in S}w_{i})\wedge
y_{s}^{\ast })$

where $w_{N}=top(\mathbf{\succcurlyeq }_{[O]}^{\circ })$ i.e. $w_{i}=x$ for
all $i\in \left \{ 1,...,k\right \} $, $w_{i}=z$ for all $i\in \left \{
k+1,...,2k\right \} $, and $w_{n}=0$.

By construction, $f(w_{N})$ is the l.u.b. of a nonempty family $\mathcal{T}$
of terms belonging to some of the following jointly exhaustive, partially
overlapping classes:

$T_{1}=\left \{ \wedge _{j\in J}v_{j}:J\text{ \ is a finite set }J\text{ and
there exists }j\in J\text{ such that }v\text{ }_{j}=0\right \} ,$

$T_{2}=\left \{ \wedge _{j\in J}v_{j}:J\text{ \ is\ a finite set }J\text{
and there exist }j,h\in J\text{ such that }v_{j}=x\text{ and }%
v_{h}=z\right
\} ,$

$T_{3}=\left \{ 
\begin{array}{c}
\wedge _{j\in J}v_{j}:\text{ }J\text{ is a finite set and there exists } \\ 
\text{a nonempty }J^{\prime }\subseteq J\text{ such that }|J^{\prime }|\leq
k<q\text{ and }v_{j}=x\text{ for all }j\in J^{\prime }%
\end{array}%
\right \} ,$

$T_{4}=\left \{ 
\begin{array}{c}
\wedge _{j\in J}v_{j}:\text{ }J\text{ is a finite set and there exists } \\ 
\text{a nonempty }J^{\prime }\subseteq J\text{ such that }|J^{\prime }|\leq
k<q\text{ and }v_{j}=z\text{ for all }j\in J^{\prime }%
\end{array}%
\right \} $.

Notice that, again, $t=\wedge _{j\in J}v_{j}=0$ for all $t\in T_{1}\cup
T_{2} $ . Moreover, by construction, $t=\wedge _{j\in J}v_{j}<x$ for all $%
t\in T_{3}$ and $t=\wedge _{j\in J}v_{j}<z$ for all $t\in T_{4}$. Since both 
$x$ and $y$ are atoms of $\mathcal{X}$, it follows that $t=\wedge _{j\in
J}v_{j}=0$ for all $t\in T_{3}\cup $ $T_{4}$ whence $f(w_{N})=0$.

If ($\beta (ii)$) obtains then take preference profile

$\mathbf{\succcurlyeq }_{[E]}^{\circ }=((\succcurlyeq _{i}^{\circ })_{i\in
\left \{ 1,...,k\right \} },(\succcurlyeq _{i}^{\circ \circ })_{i\in
\left
\{ k+1,...,2k-1\right \} },\succcurlyeq _{n}^{\circ \circ \circ })$,

and compute $f(w_{N}^{\prime })=\vee _{S\in \mathcal{F}}((\wedge _{i\in
S}w_{i}^{\prime })\wedge y_{s}^{\ast })$,

where $w_{N}^{\prime }=top(\mathbf{\succcurlyeq }_{[E]}^{\prime })$ i.e. $%
w_{i}^{\prime }=x$ for all $i\in \left \{ 1,...,k\right \} $, $w_{i}^{\prime
}=z$ for all $i\in \left \{ k+1,...,2k-1\right \} $, and $w_{n}^{\prime }=0$.

Again, $f(w_{N}^{\prime })$ is the l.u.b. of a nonempty family $\mathcal{T}$
of terms belonging to some of the following jointly exhaustive, partially
overlapping classes:

$T_{1}^{\prime }=\left \{ \wedge _{j\in J}v_{j}:J\text{ \ is a finite set }J%
\text{ and there exists }j\in J\text{ such that }v\text{ }_{j}=0\right \} ,$

$T_{2}^{\prime }=\left \{ 
\begin{array}{c}
\wedge _{j\in J}v_{j}:J\text{ \ is\ a finite set }J\text{ and there exist }
\\ 
j,h\in J\text{ such that }v_{j}=x\text{ and }v_{h}=z%
\end{array}%
\right \} ,$

$T_{3}^{\prime }=\left \{ 
\begin{array}{c}
\wedge _{j\in J}v_{j}:\text{ }J\text{ is a finite set and there exists } \\ 
\text{a nonempty }J^{\prime }\subseteq J\text{ such that }|J^{\prime }|<q%
\text{ and }v_{j}=x\text{ for all }j\in J^{\prime }%
\end{array}%
\right \} ,$

$T_{4}^{\prime }=\left \{ 
\begin{array}{c}
\wedge _{j\in J}v_{j}:\text{ }J\text{ is a finite set and there exists } \\ 
\text{a nonempty }J^{\prime }\subseteq J\text{ such that }|J^{\prime }|<q%
\text{ and }v_{j}=z\text{ for all }j\in J^{\prime }%
\end{array}%
\right \} $.

Notice that $t=\wedge _{j\in J}v_{j}=0$ for all $t\in T_{1}^{\prime }$, and
for all $t\in T_{2}^{\prime }$ as well since $x\wedge z=0$. Moreover, since $%
f(w_{N}^{\prime })=\vee _{S\in \mathcal{F}}((\wedge _{i\in S}w_{i}^{\prime
})\wedge y_{s}^{\ast })$, it also follows that $t=\wedge _{j\in J}v_{j}<x$
for all $t\in T_{3}^{\prime }\cap \mathcal{T}$ and $t=\wedge _{j\in
J}v_{j}<z $ for all $t\in T_{4}^{\prime }\cap \mathcal{T}$ . On the other
hand, $t_{2}\geqslant x$ and $t_{3}\geqslant z$ for any $t_{2}\in
T_{2}^{\prime }$ and $t_{3}\in T_{3}^{\prime }$.

It follows, again, that $\ f(w_{N}^{\prime })=0$.

Now, take $u_{N}^{\prime }\in X^{N}$ with $u_{i}^{\prime }=x\vee z$ for all $%
i\in \left \{ 1,...,n-1\right \} =N\smallsetminus \left \{ n\right \} $, and 
$u_{n}^{\prime }=0.$ By construction,

$f(u_{N}^{\prime })=\vee _{S\in \mathcal{F}}((\wedge _{i\in S}u_{i}^{\prime
})\wedge y_{s}^{\ast })$

is the l.u.b. of a nonempty family $\mathcal{T}$ of terms belonging to some
of the following jointly exhaustive, partially overlapping classes:

$T_{1}^{\prime \prime }=\left \{ \wedge _{j\in J}v_{j}:J\text{ \ is a finite
set }J\text{ and there exists }j\in J\text{ such that }v\text{ }%
_{j}=0\right
\} ,$

$T_{2}^{\prime \prime }=\left \{ 
\begin{array}{c}
\wedge _{j\in J}v_{j}:\text{ }J\text{ is a finite set and there exists} \\ 
\text{a nonempty }J^{\prime }\subseteq J\text{ such that }|J^{\prime }|<q%
\text{ and }v_{j}=x\vee z\text{ for all }j\in J^{\prime }%
\end{array}%
\right \} ,$

$T_{3}^{\prime \prime }=\left \{ 
\begin{array}{c}
\wedge _{j\in J}v_{j}:\text{ }J\text{ is a finite set and there exists } \\ 
J^{\prime }\subseteq J\text{ such that }|J^{\prime }|\geq q\text{ and }%
v_{j}=x\vee z\text{ for all }j\in J^{\prime }%
\end{array}%
\right \} $.

Observe that $t=\wedge _{j\in J}v_{j}=0$ for all $t\in T_{1}^{\prime \prime
} $. Moreover, by definition of $f$ and $q$, both $y_{s^{\prime }}^{\ast }<x$
and $y_{s^{\prime }}^{\ast }<z$ for all $s^{\prime }<q$, hence $t=\wedge
_{j\in J}v_{j}=0$ for all $t\in T_{2}^{\prime \prime }$ as well.
Furthermore, $T_{3}^{\prime \prime }\cap \mathcal{T\neq \varnothing }$ and ,
by definition of $f$ and $q$, it must be the case that for all $s\geq q$,
both $\ x\leqslant y_{s}^{\ast }$ and $z\leqslant y_{s}^{\ast }$ hold.
Therefore, $x\vee z\leqslant y_{s}^{\ast }$. It follows that $%
f(u_{N}^{\prime })=x\vee z.$

Thus, if $n=2k+1$, $f(u_{N}^{\prime })\succ _{i}f(w_{N})$ for all $i\in
N\smallsetminus \left \{ n\right \} $.

Similarly, if $n=2k,$ then $f(u_{N}^{\prime })$ $\succ _{i}f(w_{N}^{\prime
}) $ for all $i\in N\smallsetminus \left \{ n\right \} $.

Hence, $f$ is coalitionally manipulable at unimodal preference profile $%
\mathbf{\succcurlyeq }_{[O]}^{\circ }\in U_{\mathcal{X}}^{N}$ (at unimodal
preference profile $\mathbf{\succcurlyeq }_{[E]}^{\circ }\in U_{\mathcal{X}%
}^{N}$, respectively), a contradiction again, and the proof is complete.

The locally strictly unimodal case can be addressed precisely by the same
argument, provided preference profile $(\succcurlyeq ^{\ast },\succcurlyeq
^{\ast \ast },\succcurlyeq ^{\ast \ast \ast })$ is replaced by any locally
strictly unimodal preference profile $(\succcurlyeq ^{+},\succcurlyeq
^{++},\succcurlyeq ^{+++})$ such that

$\succcurlyeq ^{+}:=[x\succ ^{+}x\vee z\succ ^{+}0\succ ^{+}z\succ ^{+}w$
for all $w\in X\smallsetminus Y]$,

$\succcurlyeq ^{++}:=[z\succ ^{++}x\vee z\succ ^{++}0\succ ^{++}x\succ
^{++}w $ for all $w\in X\smallsetminus Y]$,

$\succcurlyeq ^{+++}:=[0\succ ^{+++}x\succ ^{+++}z\succ ^{+++}x\vee z\succ
^{+++}w$ for all $w\in X\smallsetminus Y]$.
\end{proof}

\section{Appendix 2: Tree automata}

Tree automata are a powerful generalization of the more widely known
sequential automata (see chpt. 2 of Ad\'{a}mek and Trnkov\'{a} (1990) for a
thorough treatment of tree automata in a categorial framework).

A \textit{finitary type }is a pair $\Sigma =(S,\alpha )$ where $S$ is a set
(whose members denote operation symbols) and $\alpha \in \mathbb{N}^{S}$ is
a function mapping $S$ into natural numbers which specifies for each $s\in S$
the corresponding (finitary) `arity' $\alpha (s)\in \mathbb{N}$ of the
corresponding operation.\footnote{%
The \textit{degree }$m=$ $m(\Sigma )$ of finitary type $\Sigma $ is the
largest `arity' of an operation denoted \ by one of its symbols i.e. $%
m(\Sigma )=\vee _{s\in S}\alpha (s)$ (if \ such `arities' are unbounded
posit $m(\Sigma )=\omega $ where $\omega =|\mathbb{N}|$). As usual, $m$ is
identified here with the set of all natural numbers smaller than $m$,
starting with $0$.} A $\Sigma $-\textit{algebra }is a pair $A=(X,\left \{
f_{s}\right \} _{s\in S})$ where $X$ is a set and, for each $s\in \Sigma ,$ $%
f_{s}:X^{\alpha (s)}\rightarrow X$ is an $\alpha (s)$-ary \textit{operation }%
on $X$. \ For any pair of $\Sigma $-algebras $A=(X,\left \{ f_{s}\right \}
_{s\in S})$, $B=$ $(X^{\prime },\left \{ f_{s}^{\prime }\right \} _{s\in S})$
a \textit{homomorphism }of $A$ into $B$ is an operation-preserving function $%
\varphi :X\rightarrow X^{\prime }$, namely for each $s\in S$ and $%
x_{1},...,x_{\alpha (s)}\in X$, $f_{s}^{\prime }(\varphi (x_{1}),...,\varphi
(x_{\alpha (s)}))=\varphi (f_{s}(x_{1},...,x_{s}))$.

A \textit{non-initial }$\Sigma $-\textit{tree automaton} is a quadruple $%
\mathcal{A=}(Q,\left \{ d_{s}\right \} _{s\in S},Y,h)$ where $Q$ is a set,
the \textit{set of states}, $d_{s}:Q^{\alpha (s)}\rightarrow Q$ is $\alpha
(s)$-ary \textit{operation }on $Q$ for any $s\in S$ , $Y$ is a set, the 
\textit{output alphabet}, and $h:Q\longrightarrow Y$ is the \textit{output
function}: thus, $\mathcal{A}$ amounts to a $\Sigma $-algebra $(Q,\left \{
d_{s}\right \} _{s\in S})$ supplemented with an output alphabet and an
output function modeling the `external' effects or observable behaviour of
the former.

A (initial) $\Sigma $-\textit{tree automaton }is a sixtuple $\mathcal{A}%
^{I,\lambda }=(Q,\left \{ d_{s}\right \} _{s\in S},Y,h,I,\lambda )$ where

$\mathcal{A}=(Q,\left \{ d_{s}\right \} _{s\in S},Y,h)$ is a non-initial $%
\Sigma $-tree automaton, $I$ is a set, the \textit{set of variables}, and $%
\lambda :I\rightarrow Q$ is the \textit{initialization function}.

For any set $I$ of variables, a \textit{finite labelled }$(\Sigma ,I)$%
\textit{-tree} is a triple $T=(P,\leqslant ,p_{0})$ such that: (i) $%
P\subseteq \Sigma \cup I$ is the \textit{finite }set of \textit{nodes}, (ii) 
$\leq $ is a \textit{partial order on }$P$\textit{\ with the tree property}
namely for any $p\in P$ the set $p\downarrow =\left\{ q\in P:q\leq p\right\} 
$ of $\leq $-predecessors of $p$ is linearly ordered i.e. is a chain, (iii) $%
p_{0}\in P$ is the \textit{root} of $\mathcal{T}$ \ i.e. the minimum of $%
(P,\leq )$, (iv) for any $p\in P$ if $p\in I$ or $p=s$ for some $s\in S$
such that $\alpha (s)=0$ then $p$ is $\leq $-maximal (or a \textit{terminal
node }of $T$); (v) for any $p\in P$ if $p=s\in \Sigma $ then $p$ is the 
\textit{lower cover} (or immediate $\leq $-predecessor) of precisely $\alpha
(s)$ nodes.

Observe that any $p\in P$ induces a finite labelled $(\Sigma ,I)$-tree $%
(p\uparrow ,\leqslant _{|p\uparrow },p)$, the \textit{sub-}$(\Sigma ,I)$%
\textit{-tree of }$T$\textit{\ with root} $p$ (where $p\uparrow =\left \{
q\in P:p\leq q\right \} $). In particular, each terminal node $p$ may be
identified with a \textit{degenerate }one-node finite labelled $(\Sigma ,I)$%
-tree $T_{p}=(\left \{ p\right \} ,=,p)$.

The $\mathcal{A}^{I,\lambda }$\textit{-initialized }version of a finite
labelled $(\Sigma ,I)$-tree $T$, denoted by $T(\mathcal{A}^{I,\lambda })$,
is obtained from $T$ by substituting state $\lambda (p)\in Q$ for each
variable $p\in P\cap I$.

The set of all finite labelled $(\Sigma ,I)$-trees is denoted $\mathcal{T}%
_{I}$ and can be naturally endowed with the structure of a $\Sigma $-algebra
by positing for any $s\in S$, $\psi _{s}:\mathcal{T}_{I}^{\alpha
(s)}\longrightarrow \mathcal{T}_{I}$ defined as follows: for each $%
T_{1},...,T_{\alpha (s)}\in \mathcal{T}_{I}$,

$\psi _{s}(T_{1},...,T_{\alpha (s)})$ is the finite labelled $(\Sigma ,I)$%
-tree having $s$ as its root, immediately followed by trees $%
T_{1},...,T_{\alpha (s)}$ themselves. Observe that each $z\in I$ can be
identified with a trivial one-node tree $t_{z\text{ }}$in $\mathcal{T}_{I}$.

Moreover, it can be easily checked that $(\mathcal{T}_{I},\left \{ \psi
_{s}\right \} _{s\in S})$ is in fact \textit{the free }$\Sigma $\textit{%
-algebra generated by} $I$, namely for each $\Sigma $-algebra $(Q,\left \{
d_{s}\right \} _{s\in S})$ and for each function $\lambda :I\rightarrow Q$
there exists a unique homomorphism $\rho :\mathcal{T}_{I}\rightarrow Q$ \ of 
$(\mathcal{T}_{I},\left \{ \psi _{s}\right \} _{s\in S})$ into $(Q,\left \{
d_{s}\right \} _{s\in S})$ extending $\lambda $\ to the entire set $\mathcal{%
T}_{I}$ of finite labelled $(\Sigma ,I)$-trees.

A $\Sigma $-tree automaton $\mathcal{A}^{I,\lambda }=(Q,\left \{
d_{s}\right
\} _{s\in S},Y,h,I,\lambda )$ acts on a finite labelled $%
(\Sigma ,I)$-tree $T $ by initializing it through $\lambda $, computing the
value at $T(\mathcal{A}^{I,\lambda })$ of the \textit{run map }of $\mathcal{A%
}^{I,\lambda }$ i.e. the unique homomorphism $\rho :\mathcal{T}%
_{I}\rightarrow Q$ \ of $(\mathcal{T}_{I},\left \{ \psi _{s}\right \} _{s\in
S})$ into $(Q,\left \{ d_{s}\right
\} _{s\in S})$ extending $\lambda $, and
taking $h(\rho ($ $T(\mathcal{A}^{I,\lambda }))$ as the output of $\Sigma $%
-tree automaton $\mathcal{A}^{I,\lambda }$ when applied to finite labelled $%
(\Sigma ,I)$-tree $T$. \ Hence, the action of $\mathcal{A}^{I,\lambda }$ on $%
\mathcal{T}_{I}$ is summarized by the \textit{behaviour map of }$\mathcal{A}%
^{I,\lambda }$, namely $\mathcal{A}^{I,\lambda }=h\circ \rho :\mathcal{T}%
_{I}\rightarrow Y$.

That computation, namely $\mathcal{A}^{I,\lambda }(T)$ can also be described
as a finite nested sequence $(T^{(i)}=(P^{(i)},\leq
^{(i)},p_{0}^{(i)})_{i=0,...k})$ of finite labelled $\Sigma $-trees denoting
the steps of a backward induction algorithm, namely

(i) $T^{(0)}=T$, $T^{(1)}=T(\mathcal{A}^{I,\lambda })$, $T^{(k)}=T_{q}=(%
\left \{ q\right \} ,=,q)$ for some $q\in Q$, and

(ii) for any $i=1,...,k-1$, $P^{(i+1)}\subseteq P^{(i)}$, $\leq
^{(i+1)}=\leq ^{(i)}\cap (P^{(i+1)}\times P^{(i+1)})$, and $T^{(i+1)}$ is
obtained from $T^{(i)}$ by replacing a non-terminal node labelled by some
operation symbol $s\in S$ (having only terminal nodes labelled $%
q_{1},...,q_{\alpha (s)}$ as immediate $\leq $-successors) with a new
terminal node labelled with state $\delta _{s}(q_{1},...,q_{\alpha (s)})$.

Notice, however, that since those trees amount to sub-$(\Sigma ,I)$-trees of 
$T$ and can therefore be identified with their roots, it follows that $%
\mathcal{A}^{I,\lambda }(T)=h(d_{p_{0}}(...(d_{s}(q_{1},...,q_{\alpha
(s)})...))$ where $s=p$ is the immediate $\leqslant $-predecessor of some
terminal node, $q_{1}=\lambda (z_{1}),...,q_{\alpha (s)}=\lambda (z_{\alpha
(s)})$ for some $z_{1},...,z_{\alpha (s)}\in I$, namely $\mathcal{A}%
^{I,\lambda }(T)$ -the \textit{behaviour of }$\mathcal{A}^{I,\lambda }$ at $%
T $- can also be equivalently written as the \textit{output}-value of the
outcome of a nested sequence of $\Sigma $-operations dictated by $T$ and
applied to $\lambda (I)$ that detail the computation steps of the \textit{%
run map of }$\mathcal{A}^{I,\lambda }$ at $T$.

\end{document}